\documentclass[journal]{IEEEtran}
\usepackage{amsmath, amsfonts, bbm}
\usepackage{array}
\usepackage{multirow}
\usepackage{textcomp}
\usepackage{url}
\usepackage{verbatim}
\usepackage{graphicx}
\def\BibTeX{{\rm B\kern-.05em{\sc i\kern-.025em b}\kern-.08em
    T\kern-.1667em\lower.7ex\hbox{E}\kern-.125emX}}
\usepackage{balance}
\usepackage{algorithm}
\usepackage{algpseudocode}
\usepackage{microtype}
\usepackage{subfigure}
\usepackage{cite}
\begin{document}
\title{Dictionary-Based Deblurring for Unpaired Data}
\author{Alok Panigrahi, Jayaprakash Katual, and  Satish~Mulleti, {\it Member, IEEE}
\thanks{\scriptsize The authors are with the Department of Electrical Engineering, Indian Institute of Technology Bombay, Mumbai, India, 400076. Email: alokiitb843@gmail.com, katualjayaprakash@gmail.com, mulleti.satish@gmail.com}%
}

\markboth{IEEE TRANSACTIONS ON IMAGE PROCESSING, October~2025}{}%

\maketitle

\begin{abstract}
Effective image deblurring typically relies on large and fully paired datasets of blurred and corresponding sharp images. However, obtaining such accurately aligned data in the real world poses a number of difficulties, limiting the effectiveness and generalizability of existing deblurring methods. To address this scarcity of data dependency, we present a novel dictionary learning based deblurring approach for jointly estimating a structured blur matrix and a high resolution image dictionary. This framework enables robust image deblurring across different degrees of data supervision. Our method is thoroughly evaluated across three distinct experimental settings: (i) full supervision involving paired data with explicit correspondence, (ii) partial supervision employing unpaired data with implicit relationships, and (iii) unsupervised learning using non-correspondence data where direct pairings are absent. Extensive experimental validation, performed on synthetically blurred subsets of the CMU-Cornell iCoseg dataset and the real-world FocusPath dataset, consistently shows that the proposed framework has superior performance compared to conventional coupled dictionary learning approaches. The results validate that our approach provides an efficient and robust solution for image deblurring in data-constrained scenarios by enabling accurate blur modeling and adaptive dictionary representation with a notably smaller number of training samples.
\end{abstract}

\begin{IEEEkeywords}
Dictionary learning, image deblurring, unpaired data, and unsupervised learning.
\end{IEEEkeywords}

\section{Introduction}
\IEEEPARstart{I}{emage} deblurring (ID) is a foundational yet persistently challenging problem in computer vision, with far-reaching implications for fields such as photography, surveillance, medical imaging, and remote sensing. Over the past decades, a vast array of methods have been proposed to address this problem, ranging from classical deconvolution and regularization techniques to more recent advances in multi-scale decomposition and deep learning. Despite these efforts, the quest for a universally interpretable and robust deblurring framework that remains effective under data scarcity remains unresolved.

State-of-the-art deep learning approaches for ID have achieved impressive results, but they are often inherently non-interpretable and heavily dependent on large amounts of paired training data to learn the low resolution (LR) to high resolution (HR) mapping \cite{Dong2014, Zhang2018, Liang2021}. In practical scenarios, however, acquiring such paired datasets is rarely feasible. As a workaround, researchers typically resort to generating synthetic paired data using manually designed degradation models \cite{Gu2019, Yue2022}. Unfortunately, these synthetic degradations are empirical and fail to capture the full diversity and complexity of real-world degradations, which in turn limits the generalizability and real-world performance of these methods \cite{Maeda2020, Wei2021}.

To overcome the limitations imposed by the scarcity of paired data, recent research has shifted towards training on unpaired datasets \cite{Maeda2020, Wei2021, Yuan2018}. These approaches often involve generating pseudo-paired data by first learning an HR-to-LR mapping, then using the synthesized pairs to train the LR-to-HR mapping. Early methods in this direction leveraged generative adversarial networks (GANs) with cycle-consistency or adversarial losses to synthesize pseudo-LR images. However, GAN-based methods are susceptible to issues such as mode collapse and require extensive fine-tuning to achieve stable results \cite{zhang2020deblurring_GAN}. More recently, diffusion models have been introduced to enhance the diversity of synthesized images \cite{Yang2023}, but this comes at the cost of content fidelity and necessitates large-scale datasets for effective training.

Despite these advances, a critical gap persists in the literature: the lack of interpretable frameworks that are robust to data scarcity, especially in the context of unpaired data scenarios frequently encountered in real-world applications. Most deep learning-based methods, while powerful, operate as black boxes, offering little insight into the underlying image structures or the nature of the learned transformations. Moreover, their data-hungry nature makes them less suitable for domains where annotated data is limited or expensive to obtain.

In contrast, classical deep learning approaches like dictionary learning (DL) offer a compelling alternative due to their inherent interpretability and data efficiency. The learned dictionary elements can be directly analyzed to reveal meaningful image features, and the approach is well-suited to scenarios with limited training data. Dictionary learning seeks to represent input data as sparse linear combinations of basic elements, learning both the elements or dictionary atoms and the sparse codes. This paradigm, which predates the deep learning era \cite{Mairal2009, Tosic2011}, was widely adopted for ID, where coupled dictionaries were used to associate LR and HR image patches via sparse coding \cite{Yang2012}, and multi-scale dictionaries were employed to model both local and non-local image priors \cite{Zhang2012}. Geometry-constrained sparse coding further enriched the representational power of these models \cite{Lu2012}.

Building on these insights, our work introduces a novel learning-based framework for interpretable ID in the case of linear blur, grounded in coupled dictionary learning. Linear blur models, which encompass a wide range of degradations including motion blur, defocus blur, and Gaussian blur, can be mathematically represented as convolution operations with known or unknown kernels \cite[Chapter~3, Section~3.4]{gonzalez2009digital}. Our approach formulates the learning of the blur matrix as the convolution of a fixed kernel over image patches as an iterative convex optimization problem. We initially address the scenario where paired HR and LR images are available, leveraging their shared sparse coefficients to jointly learn the dictionary and the blur matrix. Subsequently, we extend our framework to the more challenging unpaired setting, where only statistical properties or blur kernels are shared between LR and HR images, enabling the joint learning of HR dictionary and blur kernel even in the absence of explicit correspondences between image pairs.

During inference, an LR image is first decomposed over the learned blur and then reconstructed using the HR dictionary to yield. This process not only enhances deblurring performance but also provides interpretability by allowing direct inspection of the learned dictionaries and blur kernels, making it particularly valuable for understanding the nature of image degradations and the reconstruction process.

The main contributions of this paper are summarized as follows:
\begin{enumerate}
    \item We introduce a novel learning-based framework for interpretable image deblurring under linear blur conditions, grounded in coupled dictionary learning. Unlike conventional approaches that learn separate dictionaries, our method jointly learns the blurring matrix and the HR dictionary, resulting in a more integrated and efficient learning process that enhances deblurring performance.
    \item We propose a joint updating scheme for dictionary and blur matrix, along with a structured approach for blur matrix learning that accurately models the linear blurring process. This structured design enhances the framework's ability to handle diverse linear blurring patterns while maintaining computational efficiency.
    \item Our framework is extended to handle unpaired datasets, ensuring robust deblurring performance even in the absence of paired data. This makes our approach particularly suitable for real-world scenarios characterized by data scarcity, where obtaining paired clean and blurred images is impractical.
    \item Our method achieves effective training with only a small number of images, addressing the challenge of data scarcity. We validated our approach on synthetically generated linear Gaussian blur with the CMU-Cornell iCoseg dataset~\cite{iCoseg-database} and real-world focus blur with the FocusPath dataset~\cite{focusspath}, to demonstrate practical applicability, consistently showing substantial improvements over classical dictionary learning methods. Since deep learning methods require thousands of images and are impractical in such data-constrained scenarios, our comparisons focus on dictionary learning method approaches, which are feasible under limited data conditions.
\end{enumerate}
All things considered, our method combines interpretability, data efficiency, and reliable performance in linear image deblurring, especially in the difficult unpaired scenario.

In the sections that follow, we present the problem formulation for the given scenario. Section \ref{sec:Paired_data} discusses our approach in the context of paired data, followed by our method and results for unpaired data in Section ~\ref{sec:NC_data}. We present concluding remarks in Section~\ref{sec:conclusions}.


\section{Problem Formulation}
\label{sec:Problemformulation}
Consider a HR image $\mathbf{I}^h \in \mathbb{R}^{H_h \times W_h}$ and the corresponding LR image $\mathbf{I}^l \in \mathbb{R}^{H_l \times W_l}$, related using a linear-blur operation as
\begin{align}
\mathbf{I}^l =  \mathbf{I}^h * \mathbf{b} \label{eq:conv_operation},
\end{align}
where $\mathbf{b} \in \mathbb{R}^{k \times k}$ is a blur kernel with  $k \leq \min \left( H_{h}, W_{h} \right)$, and $*$ denotes \textit{narrow convolution}, that is without zero-padding \cite{narrorconvolution}. Due to the absence of padding, the spatial size of the output image reduces to $H_l = H_h - k + 1$ and $W_l = W_h - k + 1$, corresponding to the valid central region. Practical blur scenarios such as Gaussian or motion blur are modeled as \eqref{eq:conv_operation}.

The convolution-based representation naturally lends itself to patch-based processing, which is desired due to its computational efficiency \cite{shi2016real_patch1}. This localised framework offers a practical alternative to full-image formulations by readily accommodating parallel implementation and enabling computationally tractable solutions through dimensionality reduction \cite{lim2017enhanced_patch2}. In addition, local patches preserve structural information and spatial relationships, critical for degradation modelling more effectively than global representations \cite{dong2015image_patch3}. Hence, \eqref{eq:conv_operation} is also applicable for patches. In the rest of the paper, $\mathbf{I}^l$ and $\mathbf{I}^h$ denote either patches or images.

The operation in \eqref{eq:conv_operation} can also be written in vector-matrix form by using the fact that convolution is a linear operation. Specifically, let $\mathbf{y}^h \in \mathbb{R}^{N_h}$ and $\mathbf{y}^l \in \mathbb{R}^{N_l}$ are the respective vectorized versions of $\mathbf{I}^h$ and $\mathbf{I}^l$, with $N_{h} = H_h \times W_h$ and $N_{l} = H_l \times W_l$. Then 
\begin{align}
     \mathbf{y}^l = \mathbf{B}\, \mathbf{y}^h. \label{eq:linear-degradation}
\end{align}
The blur matrix $\mathbf{B} \in \mathbbm{R}^{N_{l} \times N_{h}}$ exhibits a structured block Toeplitz form, resulting from the inherent spatial consistency of linear blurring operations. This block Toeplitz structure allows for fast matrix-based spectral filtering and boundary condition-imposed inversion, which reduces noise amplification in ill-posed deblurring challenges \cite{hansen2006deblurring}.

In deep-learning frameworks, deblurring or learning an inverse mapping from degraded LR patches to HR counterparts amounts to estimating a parametric model $f_{\boldsymbol{\alpha}}: \mathbbm{R}^{N_{l}} \to \mathbbm{R}^{N_{h}}$ with parameters $\boldsymbol{\alpha}$. The model can be learned on a dataset 
\begin{align}
    \mathcal{D}_{\text{corr}} = \{ \mathbf{y}^{h}_{n}, \mathbf{y}^{l}_{n}\}_{n=1}^{N},
    \label{eq:paired_data}
\end{align}
with corresponding HR-LR patch pairs, by minimising the optimisation problem:
\begin{align} 
\min_{\boldsymbol{\alpha}} \frac{1}{N} \sum_{n=1}^{N} d\left( \, \mathbf{y}^{h}_{n}, \, f_{\boldsymbol{\alpha}}(\mathbf{y}^{l}_{n}) \, \right), \label{eq:opt_prob_sum}   
\end{align}
where $d(\cdot, \cdot)$ is a task-specific loss function such as pixel-wise $\ell_2$-loss, or structural similarity index measure (SSIM)-based loss, and $\boldsymbol{\alpha}$ represents the learnable parameters. Once trained, the deep-learning network $f_{\boldsymbol{\alpha}}$, realized using a DNN/CNN, is expected to deblur images/patches that are similar to the training dataset $\mathcal{D}_{\text{corr}}$. 

The success of the deep-learning models heavily relies on the availability of a large amount of labeled data $\mathcal{D}_{\text{corr}}$ - which is a practical limitation for several applications. In addition, the neural networks lack interpretability. 

To address these practical limitations, we consider an unpaired dataset 
\begin{align}
\mathcal{D}_{\text{no-corr}} = \left\{ \{\mathbf{x}^{h}_n\}_{n=1}^{N}, \, \{\mathbf{y}^{l}_m\}_{m=1}^{M} \right\}, \label{eq:unpaired_data}
\end{align}
where the relationship between these two sets of HR and LR images can vary depending on the level of supervision or the availability of correspondence information. The only information available is that each blur patch/image of the dataset is generated by a linear blur operation using a common, but unknown, blur matrix $\mathbf{B}$. We use this knowledge with the dictionary learning approach for deblurring. To this end, we categorize the problem into the following three scenarios.
\begin{enumerate}
    \item \textbf{Paired Data  (Full Correspondence):}
    In this framework, we consider a fully labeled dataset $\mathcal{D}_{\text{corr}} = \{ \mathbf{y}^{h}_{n}, \mathbf{y}^{l}_{n}\}_{n=1}^{N}$. The data will be used to formulate the proposed dictionary-learning-based method and will form the basis for the unpaired settings discussed next. 
    
    \item \textbf{Unpaired Setting (Partial Correspondence): }
    Here, we consider the dataset in \eqref{eq:unpaired_data} and assumed that there exists a subset of index pairs $\mathcal{C} \subset \{1, \dots, M\} \times \{1, \dots, N\}$ such that:
    \begin{align}
    \mathbf{y}^{l}_{m} = \mathbf{B}\, \mathbf{x}^{h}_{n}, \quad \forall \, (m, n) \in \mathcal{C}. \notag
    \end{align}
    Specifically, for a subset of data, both HR and corresponding LR patches exist; however, the correspondence is unknown.


    \item \textbf{No-correspondence Setting: }
    In this most general and challenging case, the LR dataset $\{\mathbf{y}^{l}_{m}\}_{m=1}^{M}$ and HR dataset $\{\mathbf{x}^{h}_{n}\}_{n=1}^{N}$ are completely unpaired and the number of patches in each set may differ ($M \ne N$). Specifically, for each $n \in \{1, \cdots, N\}$ we have that $\mathbf{B}\mathbf{x}^{h}_{n} \not\in \{\mathbf{y}^{l}_{m}, m = 1,\cdots, M\}$ 
    
    In this unsupervised scenario, models must learn from distributional statistics or hidden correlations between the LR and HR domains in the absence of explicit correspondence. 
\end{enumerate}

In the next section, we first discuss the proposed dictionary-learning-based deblurring approach for paired data and then discuss the frameworks for unpaired data.


\section{Deblurring via Dictionary Learning for Correspondence Data}
\label{sec:Paired_data}

Dictionary learning provides a promising alternative for deep learning based approaches by utilising sparsity-based priors \cite{kSVD}. This method represents images as sparse combinations of dictionary atoms, enabling efficient extraction of discriminative patterns. Consider a paired dataset as in \eqref{eq:paired_data}. These patches are modelled as,
\begin{align}
\mathbf{y}_{n}^{h} = \mathbf{D}^{h} \, \mathbf{c}_{n}, \text{ and } \quad \mathbf{y}_{n}^{l} = \mathbf{D}^{l}\, \mathbf{c}_{n}. \label{eq:HR_LR_DL}
\end{align}
The HR dictionary $\mathbf{D}^{h} \in \mathbb{R}^{N_{h} \times N_{c}}$, LR dictionary $\mathbf{D}^{l} \in \mathbb{R}^{N_{l} \times N_{c}}$, are comprising of $N_c$ atoms and the sparse vector $\mathbf{c}_{n} \in \mathbb{R}^{N_{c}}$ with at-most $S$ non-zero elements $\left(\|\mathbf{c}_{n}\|_{0} \leq S \right)$. Because of the correspondence patches, $\mathbf{c}_n$ is common to both representations.  Given the paired dataset $\mathcal{D}_{\text{corr}} = \{\mathbf{Y}^{h} = [\mathbf{y}_{1}^{h}, \mathbf{y}_{2}^{h}, \dots, \mathbf{y}_{N}^{h}], \mathbf{Y}^{l} = [\mathbf{y}_{1}^{l}, \mathbf{y}_{2}^{l}, \dots, \mathbf{y}_{N}^{l}]\}$, Yang et al, proposed the conventional coupled dictionary learning (CDL) algorithm \cite{Yang2012}, which solves the following problem:
\begin{align}
    \min_{\mathbf{D},\, \mathbf{C}} \;\; \| \mathbf{Y} - \mathbf{D} \mathbf{C}\|_{F}^{2} + \lambda \|\mathbf{C}\|_{1,1}, \notag \\
    \text{where, } \mathbf{Y} =     
    \begin{bmatrix}
        \mathbf{Y}^l \\
        \mathbf{Y}^h
    \end{bmatrix},
    \, \mathbf{D} = 
    \begin{bmatrix}
        \mathbf{D}^l \\
        \mathbf{D}^h
    \end{bmatrix}.    \label{eq:CDL}
\end{align}
Here, $\lambda$ is the sparsity regularization parameter, $\|\mathbf{A}\|_{F}$ is the Frobenius norm of a matrix $\mathbf{A}$, and $\|\mathbf{C}\|_{1,1} = \sum_{j=1}^{N} \|\mathbf{c}_j\|_1$ denotes the sum of $\ell_{1}$-norms across columns of $\mathbf{C} \in \mathbbm{R}^{N_{c} \times N}$. The key insight is that corresponding HR-LR patches share an identical sparse code matrix $\mathbf{C}$. However, the correspondence will be lost when we consider unpaired data. To address this issue, we propose an alternative approach where the LR and HR parts of the cost functions, that is, $\|\mathbf{Y}^l - \mathbf{D}^l\mathbf{C}\|_F^2$ and $\|\mathbf{Y}^h - \mathbf{D}^h\mathbf{C}\|_F^2$, respectively, are tied by the HR dictionary $\mathbf{D}^h$ in addition to $\mathbf{C}$. This step, though, looks trivial, will play a huge role in the non-correspondence case, as will be discussed in the following section. Before that, we would show the effectiveness of our modification for the corresponding/paired

In our adaptation, we consider the linear degradation model in \eqref{eq:linear-degradation}, where $\mathbf{y}_{n}^{l} = \mathbf{B} \mathbf{y}_{n}^{h}$ with $\mathbf{B}$ being the structured blur matrix, the LR patches must satisfy: 
\begin{align} 
\mathbf{Y}^{l} = \mathbf{BD}^{h} \mathbf{C}, \, \text{and hence} \ \mathbf{D}^{l} = \mathbf{BD}^{h}. \label{eq:LR_DL} 
\end{align}
The coupling $\mathbf{D}^{l} = \mathbf{BD}^{h}$ ensures the learned dictionaries respect the degradation physics while maintaining sparse representation with a shared dictionary for HR-LR patches. The dictionary learning problem with the coupling is given as 
\begin{align}
\min_{\mathbf{B}, \mathbf{D}^{h}, \mathbf{C}} \, \left\| \mathbf{Y}^{h} - \mathbf{D}^{h}\mathbf{C} \right\|_{F}^{2} + \left\| \mathbf{Y}^{l} - \mathbf{BD}^{h}\mathbf{C} \right\|_{F}^{2} + \lambda  \| \mathbf{C} \|_{1,1}.
\label{eq:Corespondence-Opt-Poblem}
\end{align}
This joint optimization framework allows simultaneous learning of the degradation model and dictionary representations. Importantly, as discussed earlier, the two cost functions are tied by the sparsity matrix $\mathbf{C}$ and the HR dictionary $\mathbf{D}^h$.
 
\subsection{Solving the Joint Optimization}
\label{subsec:SolvingJointOpt}
The problem in \eqref{eq:Corespondence-Opt-Poblem} is non-convex but can be decomposed into three tractable convex subproblems. These subproblems are alternately optimized while fixing other variables, as discussed next.

\subsubsection{Sparse code recovery}
For a fixed dictionary $\mathbf{D}^{h}$ and blur matrix $\mathbf{B}$, the following least absolute shrinkage and selection operator (LASSO) optimization problem is solved for $\mathbf{C}$:
\begin{align}
\min_{\mathbf{C}} \, \left\| \mathbf{Y}^{h} - \mathbf{D}^{h}\mathbf{C} \right\|_{F}^{2} + \lambda \| \mathbf{C} \|_{1,1}.
\label{eq:LASSO-Paired}
\end{align}
Different iterative algorithms like orthogonal matching pursuit (OMP) \cite{OMP}, iterative shrinkage/thresholding algorithm (ISTA) \cite{ISTA}, and fast iterative shrinkage/thresholding algorithm (FISTA) \cite{FISTA} can be used to solve for the sparse codes. In this work, FISTA is used for solving \eqref{eq:LASSO-Paired}.

\subsubsection{Dictionary update}
Next, the dictionary $\mathbf{D}^h$ is updated with HR image patches. With learned sparse codes $\mathbf{C}$, the dictionary is updated by solving the following optimization problem:
\begin{align}
\label{eq:D-update}
\min_{\mathbf{D}^{h}} \, \left\| \mathbf{Y}^{h} - \mathbf{D}^{h}\mathbf{C} \right\|_{F}^{2}.
\end{align}

This step updates each dictionary atom one by one, as in the K-singular value decomposition (K-SVD) algorithm \cite{kSVD}.

\subsubsection{Blur matrix estimation (BME)}
\label{subsubsec: BME}
Two approaches for blur estimation are considered while minimizing the concerned optimization problem
\begin{align}
\label{eq:B-update}
\min_{\mathbf{B}} \left\| \mathbf{Y}^{l} - \mathbf{BD}^{h}\mathbf{C} \right\|_{F}^{2}.
\end{align}

The first BME approach is a general recovery or BME-GR that uses a Moore–Penrose pseudo-inverse and estimates blur as
\begin{align}
\label{eq: Pseudo B}
\mathbf{B} = \mathbf{Y}^{l}(\mathbf{D}^{h}\mathbf{C})^{+},
\end{align}
where $\mathbf{A}^{+}$ denotes Moore–Penrose pseudo-inverse of $\mathbf{A}$. Though the approach is simpler, it does not result in a block-Toeplitz structured $\mathbf{B}$.

The next BME approach with structured recovery (BME-SR) uses the block-Toeplitz information of $\mathbf{B}$. Specifically, $\mathbf{B}$ has $k^2$ distinct values at known locations and it can be expressed as linear combination of $k^2$ basis matrices $\mathbf{M}_1, \mathbf{M}_2, \dots, \mathbf{M}_{k^2}$. Here, each matrix $\mathbf{M}_i$ is a binary matrix of size $N_l \times N_h$, with 1's at the positions corresponding to one of the distinct values in $\mathbf{B}$ at the known locations and 0's elsewhere. Let $\{\theta_1, \theta_2, \dots, \theta_{k^2}\}$ be the distinct values of $\mathbf{B}$, then
\begin{align}
\label{eq:B-factorization}
\mathbf{B} = \sum_{i=1}^{k^2} \theta_i \mathbf{M}_i.
\end{align}
Using \eqref{eq:B-factorization}, the optimization problem in \eqref{eq:B-update} is updated to solve for all $\theta$s as,
\begin{align}
  \min_{\{\theta_{i}\}_{i=1}^{k^2} } \ \left\| \mathbf{Y}^{l} - \sum_{i=1}^{k^2} \theta_{i}\mathbf{M}_{i}\mathbf{D}^{h}\mathbf{C} \right\|_{F}^{2}.  \label{eq:th-update}
\end{align}
Gradient descent algorithm, like the Adam Optimizer, can be used for solving problem \eqref{eq:th-update} with known $\mathbf{Y}^{l}, \ \mathbf{M}_{i}, \ \mathbf{D}^{h},$ and $\mathbf{C}$. These learned $\theta_i$s and their known locations are used for the reconstruction of the $\mathbf{B}$ matrix.

After completing the three–stage joint optimization procedure, the blur matrix $\mathbf{B}$ and the high–resolution dictionary $\mathbf{D}^{h}$ are obtained. At inference time, we compute the low-resolution dictionary, $\mathbf{D}^{l} = \mathbf{B}\mathbf{D}^{h}$.
Given a blurred image, overlapping patches are extracted and the sparse coefficient matrix $\mathbf{C}$ is estimated with the FISTA using $\mathbf{D}^{l}$. The clean patches are then recovered as $\mathbf{D}^{h}\mathbf{C}$ and aggregated to form the final restored image. The proposed method is tested on existing benchmark datasets, and the results are discussed in the following sections.

\subsection{Experiment and Results} 
To rigorously evaluate the performance of our proposed methods, we utilized four subsets from the CMU-Cornell iCoseg database~\cite{iCoseg-database}. Specifically, the four iCoseg subsets employed were \emph{Goose-Riverside}, \emph{Alaskan Brown Bear}, \emph{Monks-LAO}, and \emph{Kite-kitekid}, which comprise 19, 10, 31, and 17 images, respectively. From each subset, 7 images were selected to generate training patches for dictionary learning, and the remaining images were used for testing. Each source image in both the training and testing sets was synthetically degraded by applying Gaussian blurring with various kernel sizes and standard deviations, specifically for $k \in \{7, 9, 11, 13\}$ and $\sigma \in \{1.2, 2, 2.5\}$.

\begin{figure*}[!t]
\begin{center}
\begin{tabular}{ccccc}
\subfigure[HR]{\includegraphics[width=1.25 in]{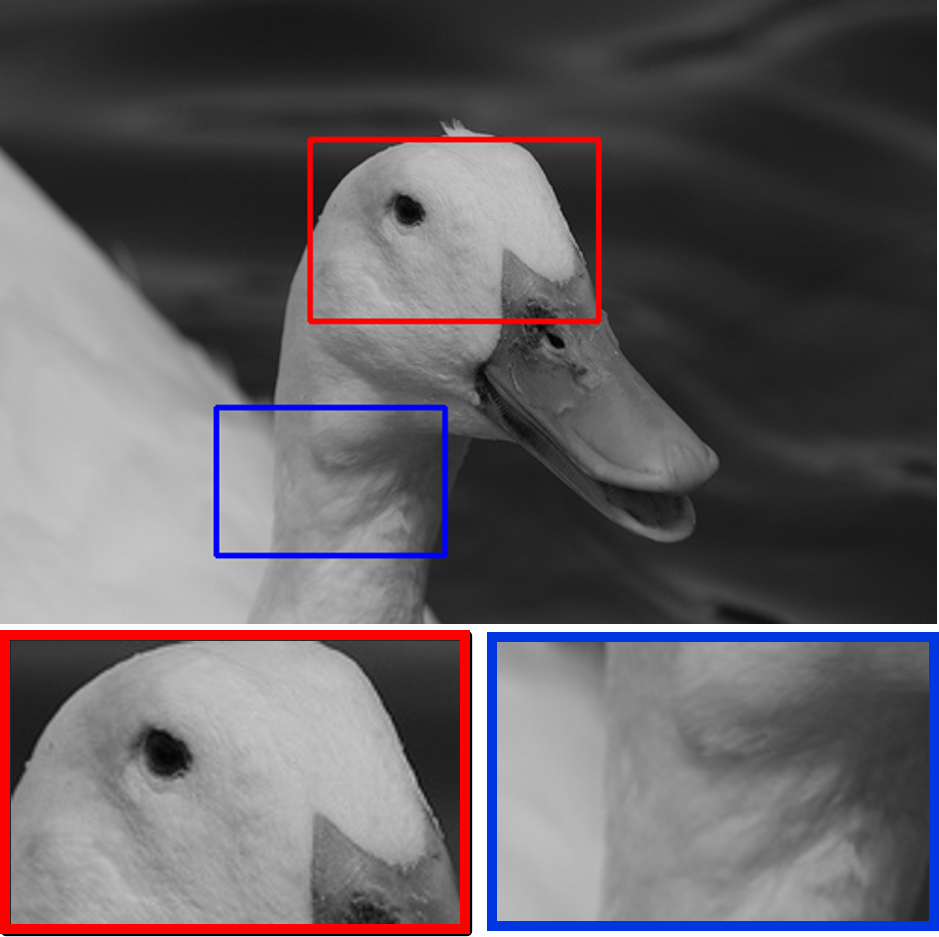}}  &
\subfigure[LR]{\includegraphics[width=1.25 in]{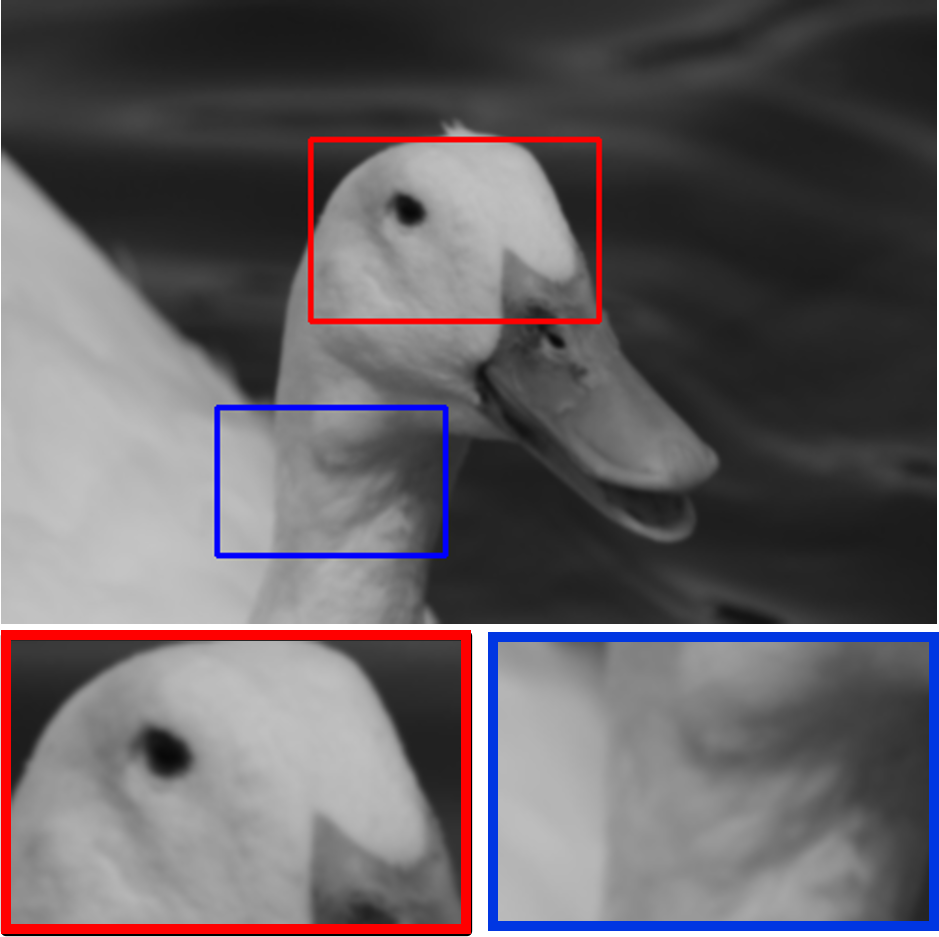}}   &
\subfigure[BME-GR]{\includegraphics[width=1.25 in]{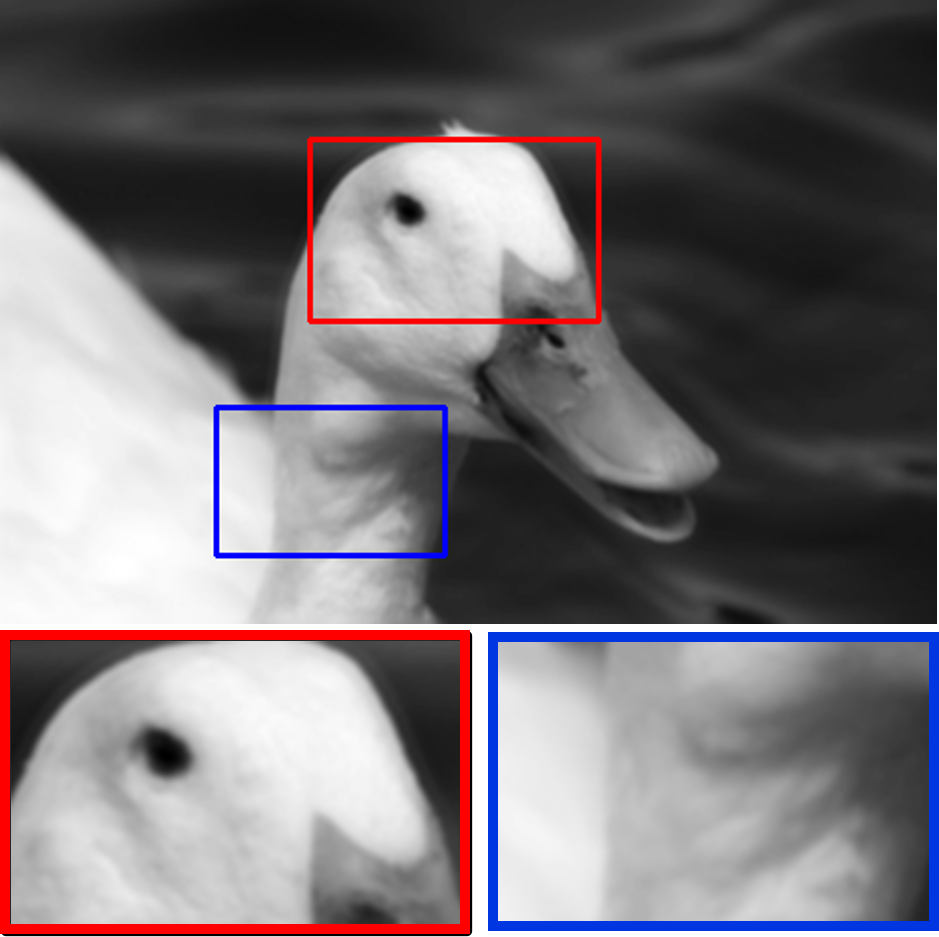}}   &
\subfigure[BME-SR]{\includegraphics[width=1.25 in]{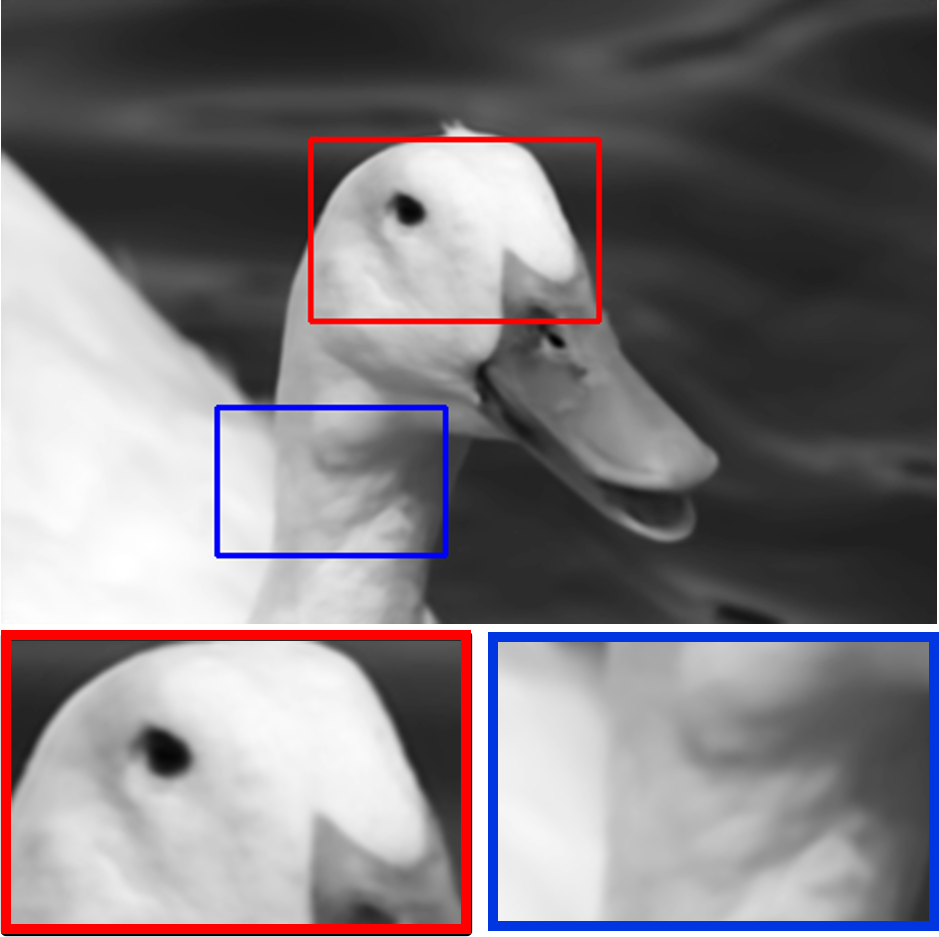}}   &
\subfigure[CDL]{\includegraphics[width=1.25 in]{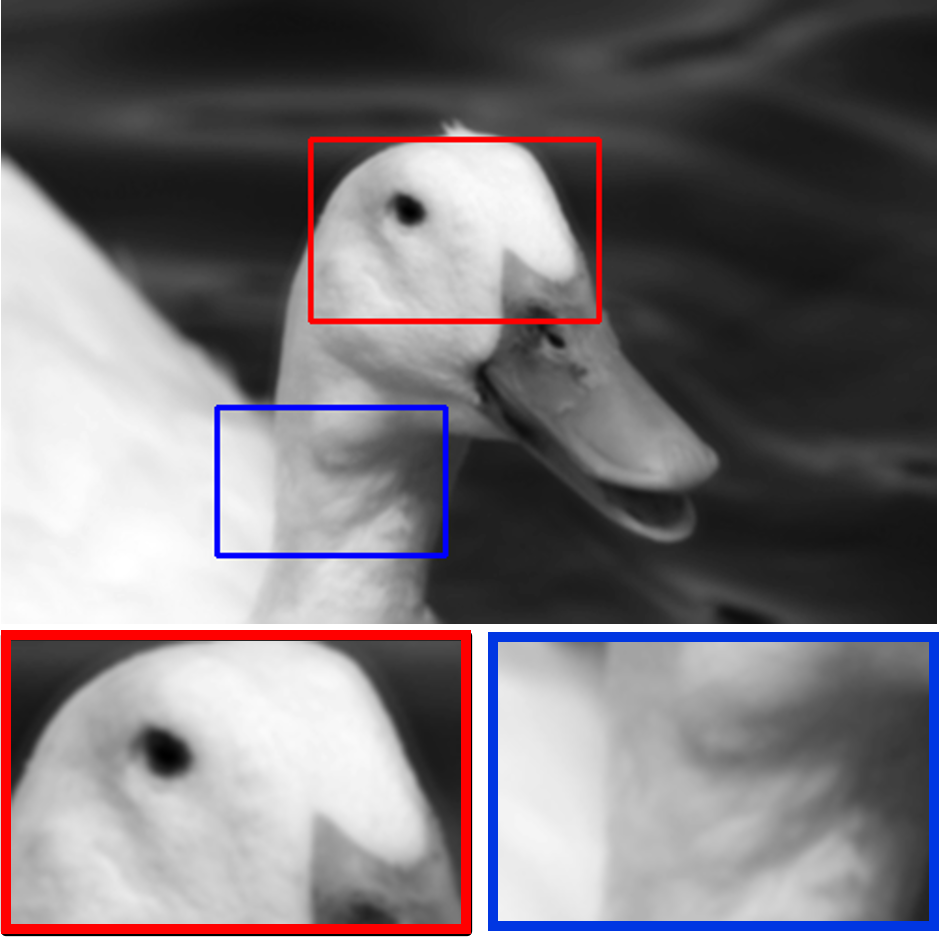}}  \\
\subfigure[HR]{\includegraphics[width=1.25 in]{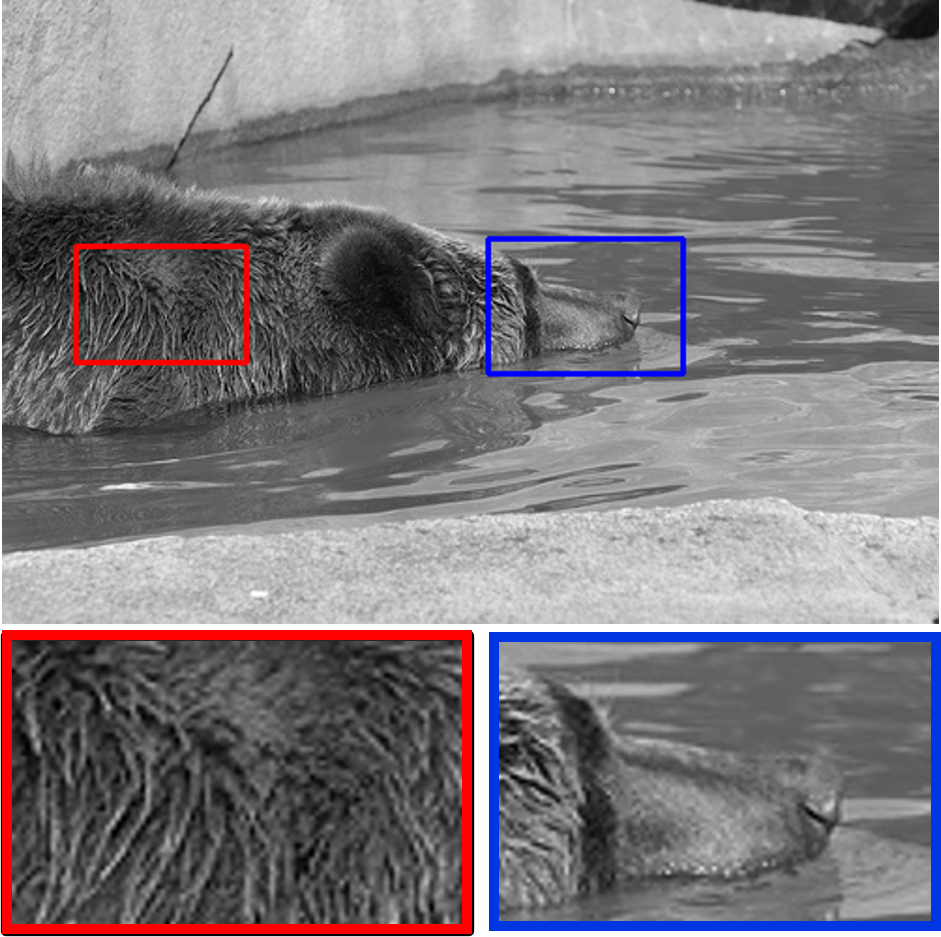}}  &
\subfigure[LR]{\includegraphics[width=1.25 in]{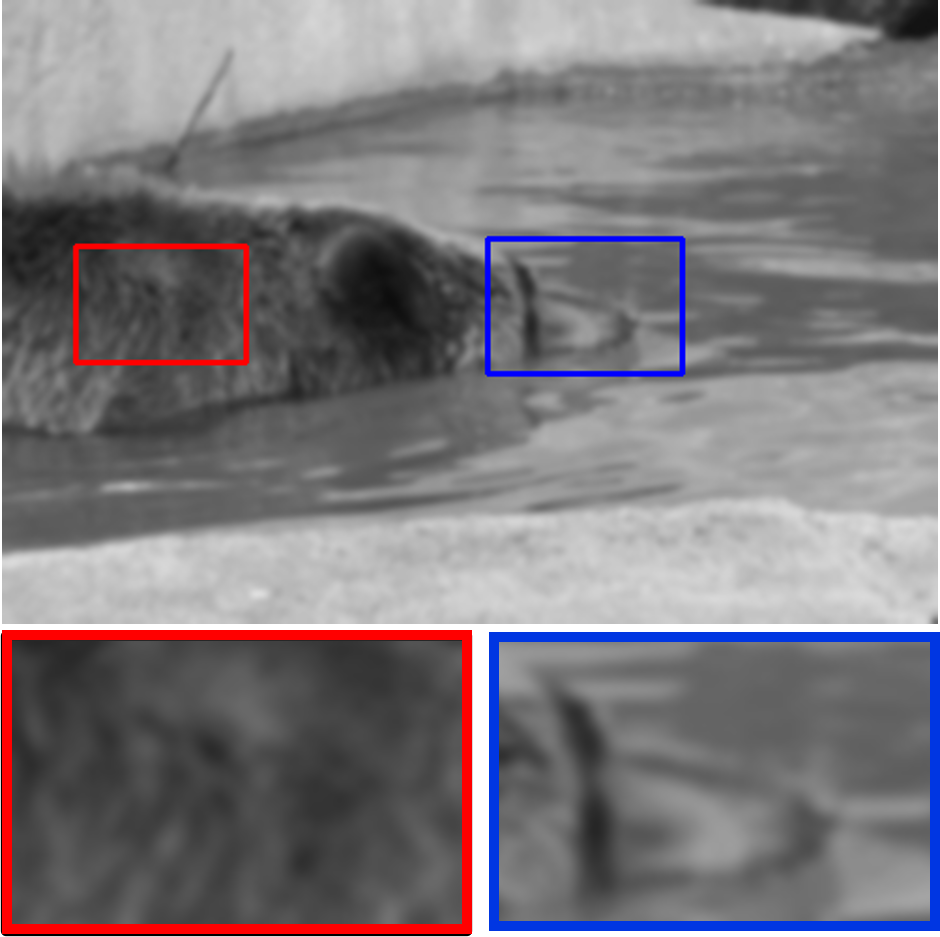}}   &
\subfigure[BME-GR]{\includegraphics[width=1.25 in]{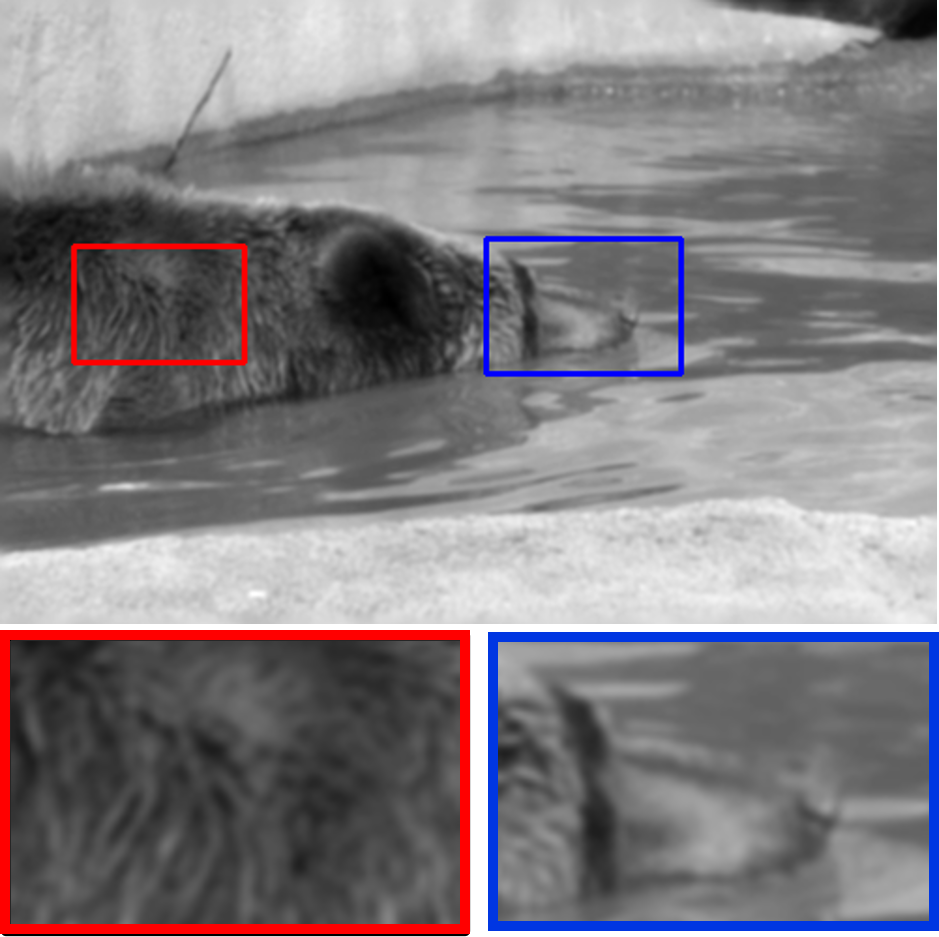}}   &
\subfigure[BME-SR]{\includegraphics[width=1.25 in]{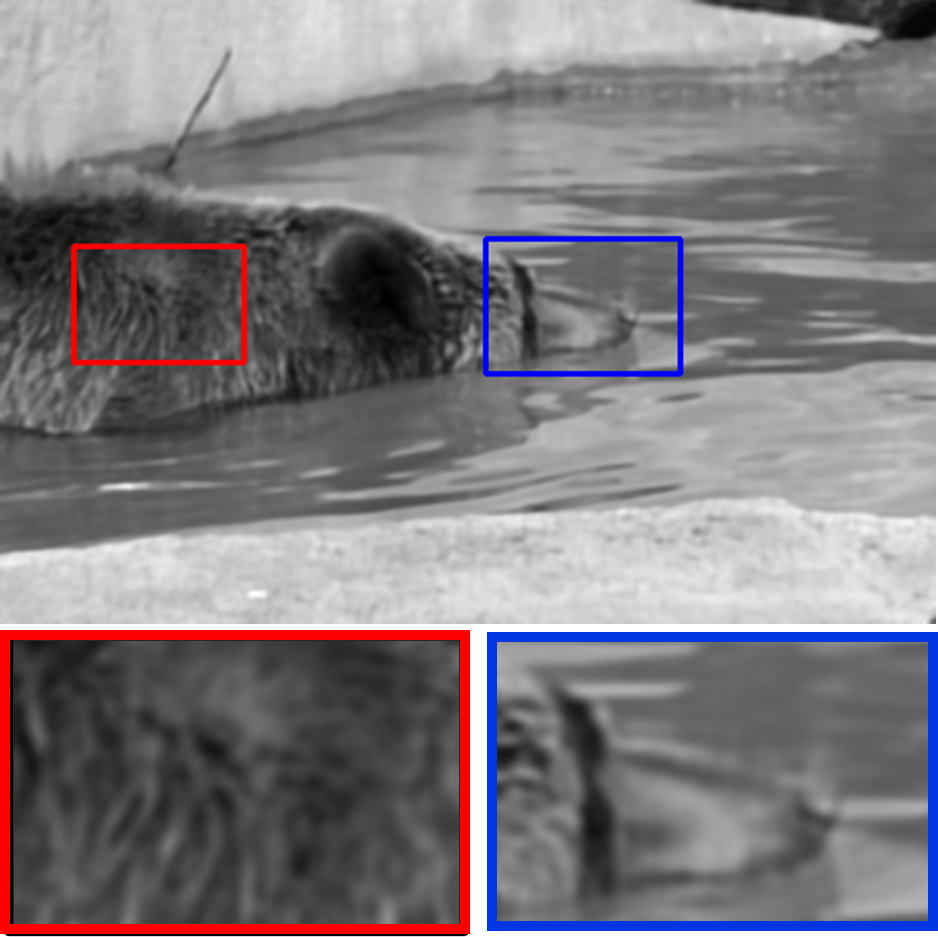}}   &
\subfigure[CDL]{\includegraphics[width=1.25 in]{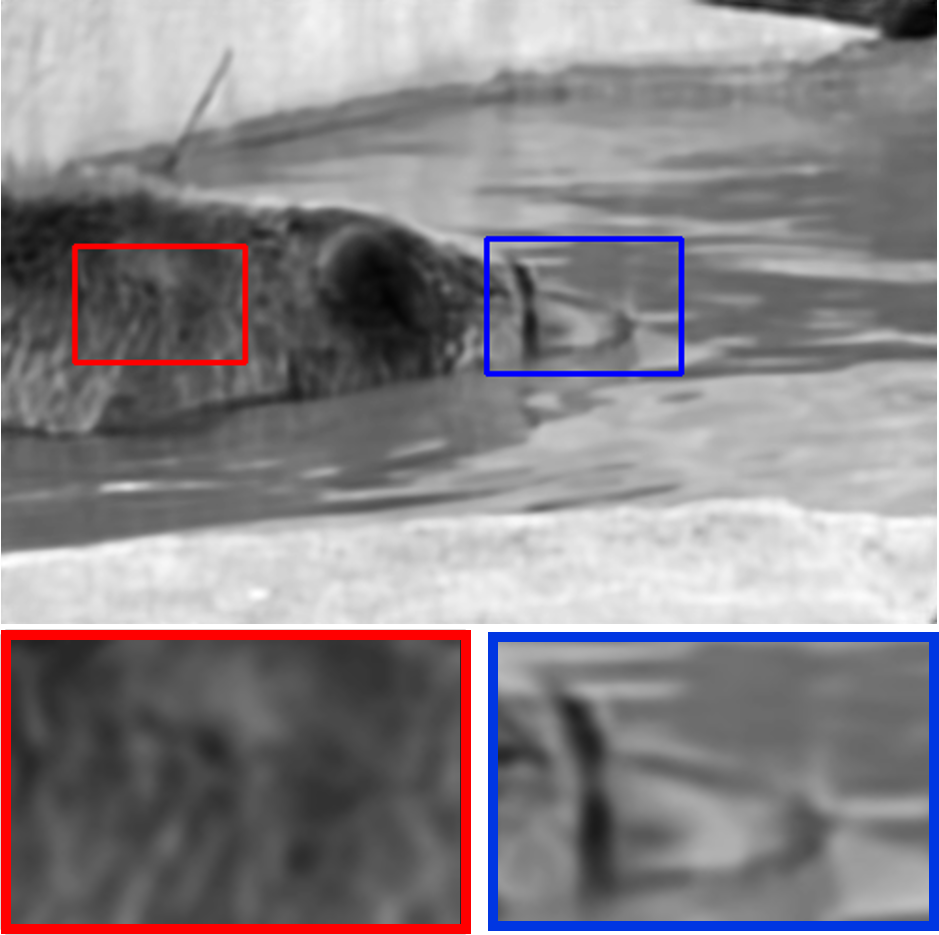}}  \\
\subfigure[HR]{\includegraphics[width=1.25 in]{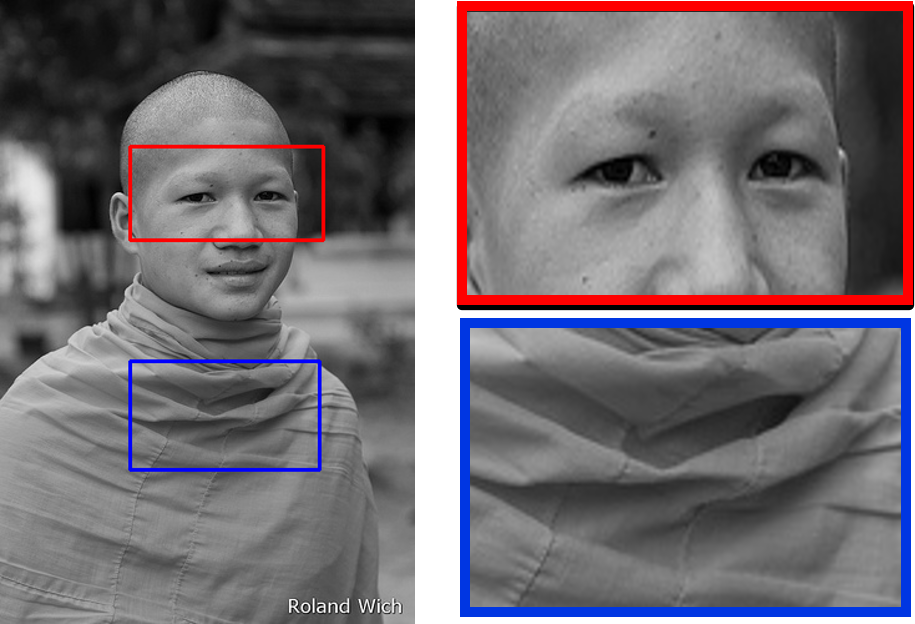}} &
\subfigure[LR]{\includegraphics[width=1.25 in]{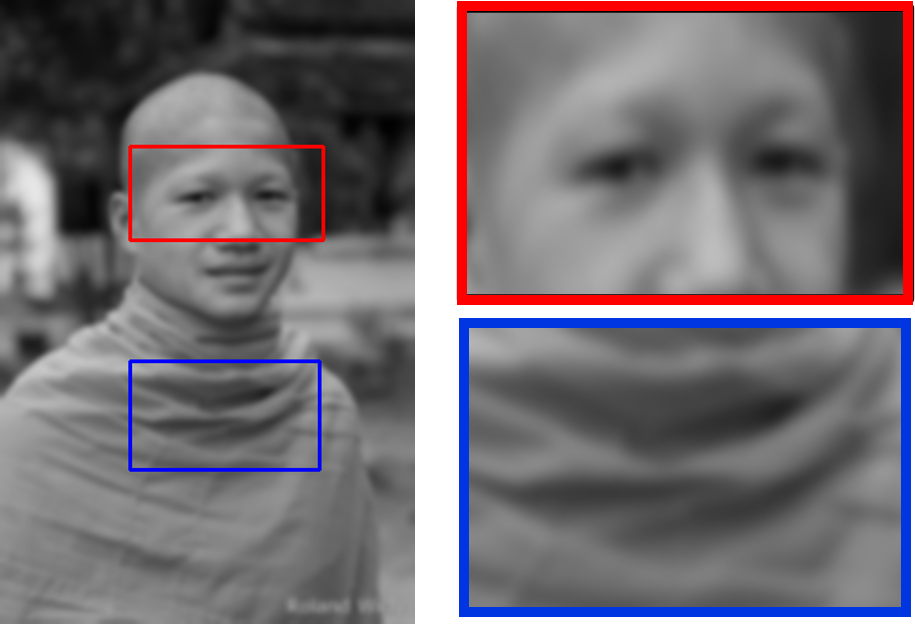}} &
\subfigure[BME-GR]{\includegraphics[width=1.25 in]{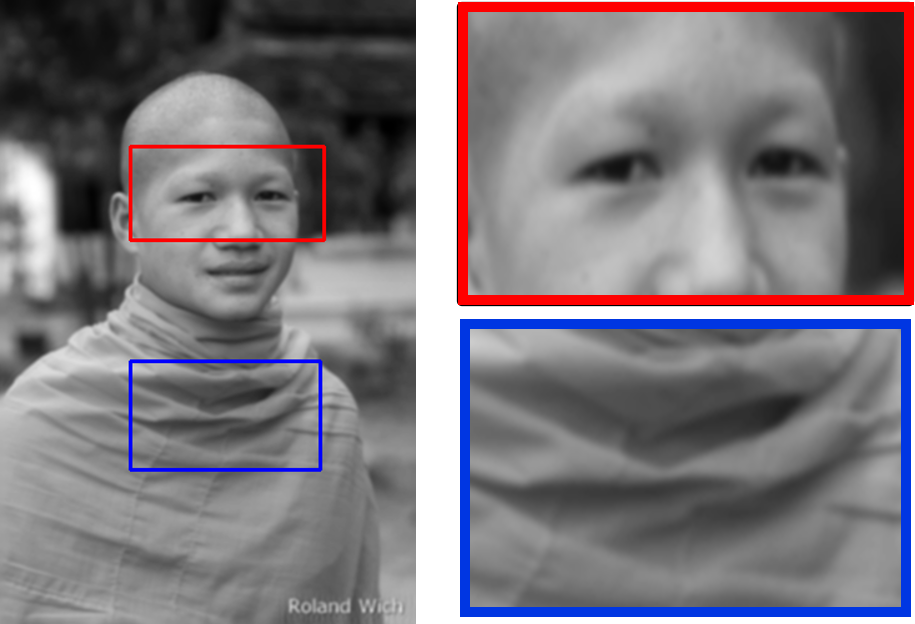}} &
\subfigure[BME-SR]{\includegraphics[width=1.25 in]{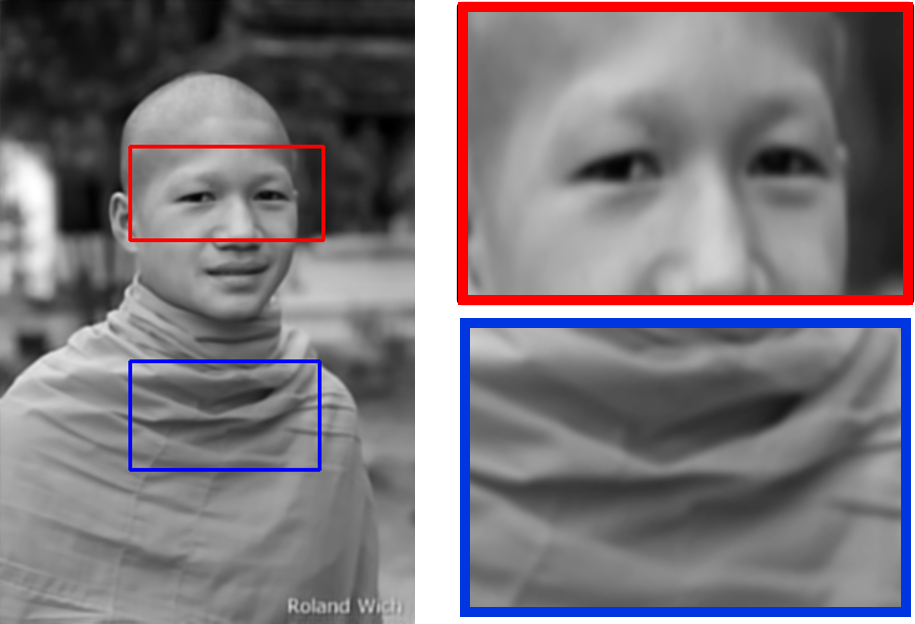}} &
\subfigure[CDL]{\includegraphics[width=1.25 in]{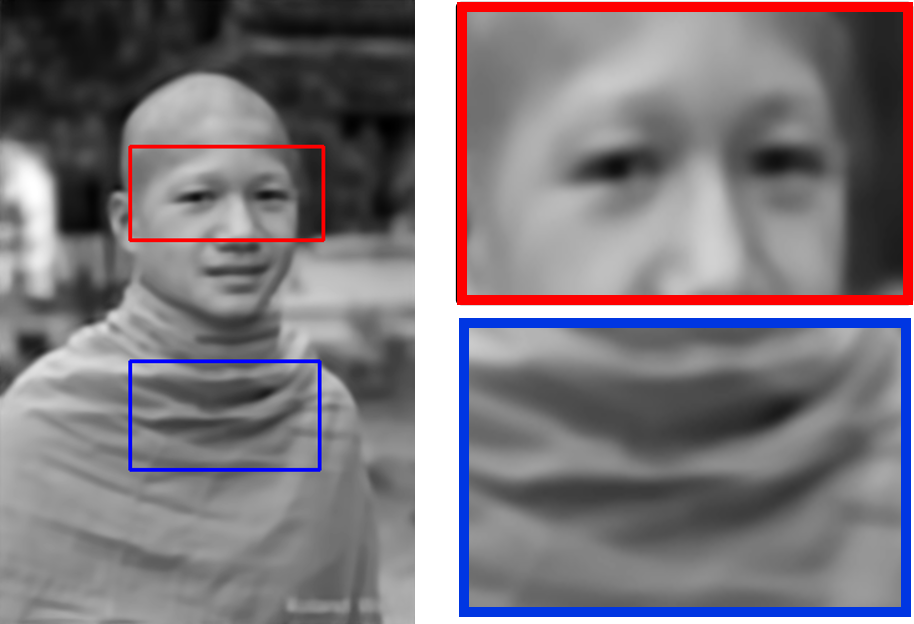}}
\end{tabular}
\caption{Visual comparison of different methods on the Gaussian dataset. (a–e) Results for $I_{1}$ (blur: $k=7$, $\sigma=1.2$); (f–j) $I_{2}$ (blur: $k=9$, $\sigma=2.0$); (k–o) $I_{3}$ (blur: $k=11$, $\sigma=2.5$). Columns show: (a, f, k) HR ground truth, (b, g, l) LR input, (c, h, m) BME-GR, (d, i, n) BME-SR, and (e, j, o) CDL results.}
\label{fig:Paired_result}
\end{center}
\end{figure*}

For algorithm training, we randomly extracted $20{,}000$ pairs of patches from the training images, each of size $15 \times 15$ pixels, from the clear and corresponding blurred versions. As previously described, we implemented and evaluated two variants of our algorithm: BME-GR and BME-SR. The hyperparameter configuration yielding the highest peak signal-to-noise ratio (PSNR) was selected for the final evaluation. The best performance was achieved with a dictionary size $N_{c} = 400$ atoms. For the FISTA algorithm, the regularization parameter $\lambda$ was set to 0.02 for the \emph{Goose-Riverside} and \emph{Monks-LAO} subsets, and 0.2 for the \emph{Alaskan Brown Bear subset}. A fixed learning rate of 0.05 was used with the Adam optimizer during gradient descent. For each case, the joint optimization was run for 5000 iterations. A high non-sparsity level was enforced to ensure that each training sample was accurately approximated using a pair of dictionaries.

\begin{figure}[t!] 
    \centering   
    \includegraphics[width=8cm]{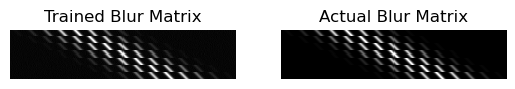}
    \caption{B Matrix obtained after solving the joint optimization}
    \label{fig:B_matrix_comparison}
\end{figure}

For qualitative analysis, we further defined three test images as $I_1$, $I_2$, and $I_3$, as shown in Fig.~\ref{fig:Paired_result} panels (a), (f) , and (k) respectively. Their blurred counter parts were generated using Gaussian kernels with parameters, $(k=7, \sigma = 1.2)$ for $I_1$, $(k=9, \sigma = 2)$ for $I_2$ and $(k=11, \sigma = 2.5)$ for $I_3$, as shown in panels (b), (g) , and (l) respectively.
Figure~\ref{fig:B_matrix_comparison} illustrates a comparison between the ground truth blur matrix $\mathbf{B}$ and the estimated blur matrix obtained using the BME-SR algorithm after training on the \emph{Goose-Riverside} dataset blurred using a Gaussian blur. The Frobenius norm error, expressed in decibels (dB), is $-30.43$~dB, indicating a high degree of similarity between the estimated and ground truth blur matrices.

We further compared our approach against the CDL algorithm \cite{Yang2012}. For a fair comparison, the optimal hyperparameters, such as the number of atoms and sparsity level, were selected for CDL by maximizing the quality of the deblurred results.

Figures~\ref{fig:Paired_result} present qualitative deblurring results for three representative test images for different $k$ and $\sigma$ values. As illustrated, our algorithms consistently produced sharper and more visually appealing reconstructions compared to CDL. For example, in  $I_{2}$, first patch, the eyes of the monk are completely recovered using our BME-GR and BME-SR methods compared to LR, also producing sharper results than the CDL method.


Quantitative results, summarized in Table~\ref{tab:paired_metrics}, report the PSNR and SSIM across varying kernel sizes and standard deviations, which further substantiates the superiority of our methods over the CDL method, with BME-SR performing slightly better than BME-GR in most cases. In particular, BME-SR demonstrates consistent performance advantages across different blur configurations, achieving PSNR improvements in the range of 1.5 to 3.5 dB compared to CDL. While BME-GR excels with smaller kernel sizes ($\sigma = 1.2$, $k = 7$), BME-SR exhibits superior robustness as the blur strength increases. Notably, the SSIM metrics follow a similar pattern, with our proposed approaches consistently outperforming CDL across most test images and blur configurations, indicating better preservation of structural information in the restored images. 

In the general scenario, where the blur kernel size $k$ was unknown, we trained for $\mathbf{B}$ and $\mathbf{D}^{h}$ using a range of kernel sizes for a given image set, and subsequently validated performance for different $k$ values using the BME-GR method. Table~\ref{tab:diff_k_paired} reports results for kernel sizes $k \in \{13, 11, 9, 7\}$ (assuming an upper bound of $17$) under different Gaussian blur. For each image, the model trained with the kernel size yielding the best validation performance was selected for deblurring. This strategy ensured that our approach remained robust and adaptable to diverse blur conditions, thereby enhancing its practical applicability.

\begin{figure*}[!t]
\begin{center}
\begin{tabular}{ccccc}
\subfigure[LR]{\includegraphics[width=1.25 in]{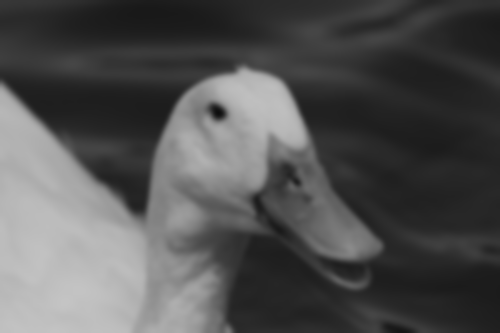}}  &
\subfigure[BME-GR ($k$ = 7)]{\includegraphics[width=1.25 in]{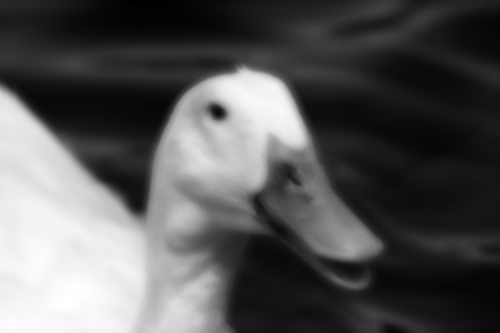}}   &
\subfigure[BME-GR ($k$ = 9)]{\includegraphics[width=1.25 in]{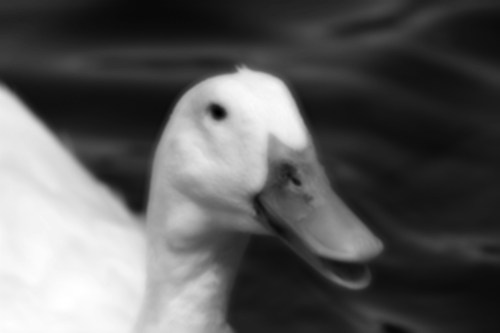}}   &
\subfigure[\textbf{BME-GR ($k$ = 11)}]{\includegraphics[width=1.25 in]{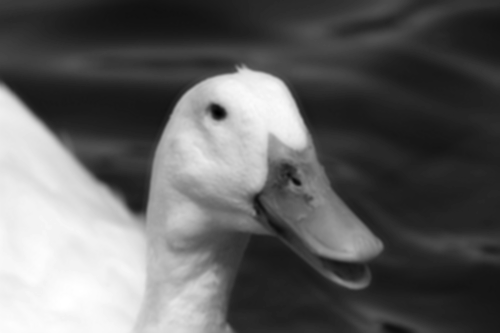}}   &
\subfigure[BME-GR ($k$ = 13)]{\includegraphics[width=1.25 in]{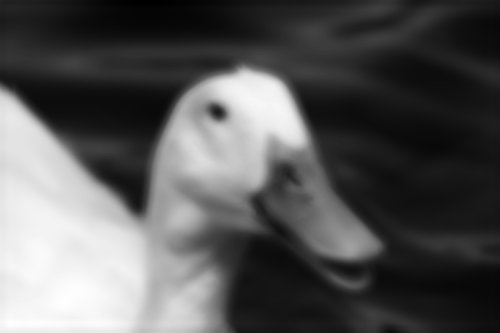}} 
\end{tabular}
\caption{Cross-validation results for BME-GR on $I_{1}$. (a) LR input; (b-e) BME-GR reconstructions with $k = 7, 9, 11, 13$ respectively.}
\label{fig:Result_p_cross_val}
\end{center}
\end{figure*}

Figure~\ref{fig:Result_p_cross_val} shows BME-GR results for $I_{1}$ trained with different kernels, with $k = 11$ giving the best results. This can be qualitatively observed from the figure and quantitatively supported by the PSNR/SSIM values in Table~\ref{tab:diff_k_paired}.

\renewcommand{\arraystretch}{1.3}
\begin{table}[!t]
\centering
\caption{Results for different $k$ and $\sigma$. (PSNR in dB)} \small
\begin{tabular}{|c|c|c|c|c|c|c|}
\hline
\multicolumn{1}{|c|}{\multirow{3}{*}{Image}} & \multicolumn{2}{c|}{BME-GR} & \multicolumn{2}{c|}{BME-SR} & \multicolumn{2}{c|}{CDL} \\  \cline{2-7}
\multicolumn{1}{|c|}{}  & PSNR  & SSIM  & PSNR  & SSIM  & PSNR  & SSIM  \\ \hline
\multicolumn{7}{|c|}{$\sigma = 1.2$, $k = 7$} \\ \hline
$I_{1}$ & \textbf{34.67} & 0.92 & 33.69 & \textbf{0.93} & 32.67 & 0.92 \\ \hline
$I_{2}$ & 28.69 & 0.88 & \textbf{29.99} & \textbf{0.91} & 28.76 & 0.89 \\ \hline
$I_{3}$ & \textbf{24.89} & 0.60 & 24.63 & \textbf{0.64} & 21.84 & 0.63 \\ \hline
$I_{4}$ & \textbf{28.98} & \textbf{0.72} & 28.36 & 0.71 & 25.48 & 0.71 \\ \hline
 \multicolumn{7}{|c|}{$\sigma = 2$, $k = 9$} \\ \hline
$I_{1}$ & 32.65 & 0.91 & \textbf{34.38} & \textbf{0.94} & 32.57 & 0.92 \\ \hline
$I_{2}$ & 28.23 & 0.86 & \textbf{29.83} & \textbf{0.92} & 28.46 & 0.90 \\ \hline
$I_{3}$ & 22.93 & 0.56 & \textbf{24.98} & \textbf{0.63} & 21.44 & 0.61 \\ \hline
$I_{4}$ & \textbf{29.01} & 0.70 & 28.96 & \textbf{0.71} & 25.01 & 0.69 \\ \hline
\multicolumn{7}{|c|}{$\sigma = 2.5$, $k = 11$} \\ \hline
$I_{1}$ & 32.06 & 0.90 & \textbf{34.31} & \textbf{0.94} & 33.72 & 0.92 \\ \hline
$I_{2}$ & 28.86 & 0.85 & \textbf{29.86} & \textbf{0.91} & 28.36 & 0.89 \\ \hline
$I_{3}$ & 22.72 & 0.54 & \textbf{24.88} & 0.61 & 23.94 & \textbf{0.63} \\ \hline
$I_{4}$ & \textbf{28.81} & 0.64 & 28.36 & \textbf{0.66} & 25.81 & 0.64 \\ \hline
\end{tabular}\
\label{tab:paired_metrics}
\end{table}
\renewcommand{\arraystretch}{1}

\begin{table}[!t]
\centering
\caption{PSNR and SSIM values for different $k$ values}\small
\begin{tabular}{|c|c|c|c|c|c|}
\hline
Image & Metric & \multicolumn{4}{c|}{$k$ values} \\
\cline{3-6}
& & $k=13$ & $k=11$ & $k=9$ & $k=7$ \\
\hline
\multirow{2}{*}{$I_{1}$} & PSNR & 29.62 & 29.21 & 30.97 & \textbf{32.48} \\
& SSIM & 0.83 & 0.85 & 0.87 & \textbf{0.90} \\
\hline
\multirow{2}{*}{$I_{2}$} & PSNR & 26.71 & \textbf{27.81} & 26.25 & 25.88 \\
& SSIM & 0.82 & \textbf{0.85} & 0.81 & 0.79 \\
\hline
\multirow{2}{*}{$I_{3}$} & PSNR & \textbf{22.89} & 21.10 & 22.67 & 22.15 \\
& SSIM & \textbf{0.56} & 0.53 & 0.55 & 0.54 \\
\hline
\multirow{2}{*}{$I_{4}$} & PSNR & 24.13 & 24.98 & 25.77 & \textbf{26.42} \\
& SSIM & 0.74 & 0.76 & 0.78 & \textbf{0.80} \\
\hline
\end{tabular}
\label{tab:diff_k_paired}
\end{table}

A fundamental advantage of the proposed dictionary learning-based methodology is its ability to achieve superior deblurring performance with substantially reduced training data requirements. Our method can be effectively trained with only 6--7 images. In contrast, deep learning methods typically require thousands of images for effective training, making them impractical in such data-constrained scenarios \cite{alzubaidi2023survey}. Therefore, our comparisons focus on classical dictionary learning methods, which are feasible under limited data conditions.

To further validate the practical applicability of our method, we conducted comprehensive evaluations on the FocusPath dataset~\cite{focusspath}. This dataset contains Whole Slide Images (WSIs) captured at 16 discrete focal depths, or $z$-levels. The images from the central focal planes, specifically $z=8$ and $z=9$, are the sharpest (in-focus) and serve as our HR ground truth references. As the focal plane moves away from this central in-focus region down towards $z=1$ or up towards $z=16$ the images become progressively more blurred due to being out of focus. This natural optical degradation can be effectively modeled as a Gaussian blur where the standard deviation, $\sigma$, increases with the distance from the HR planes, reaching maximum severity at the extremes ($z=1$ and $z=16$). 

\begin{figure*}[!t]
\begin{center}
\begin{tabular}{ccccccc}
\subfigure[HR]{\includegraphics[width=0.84 in]{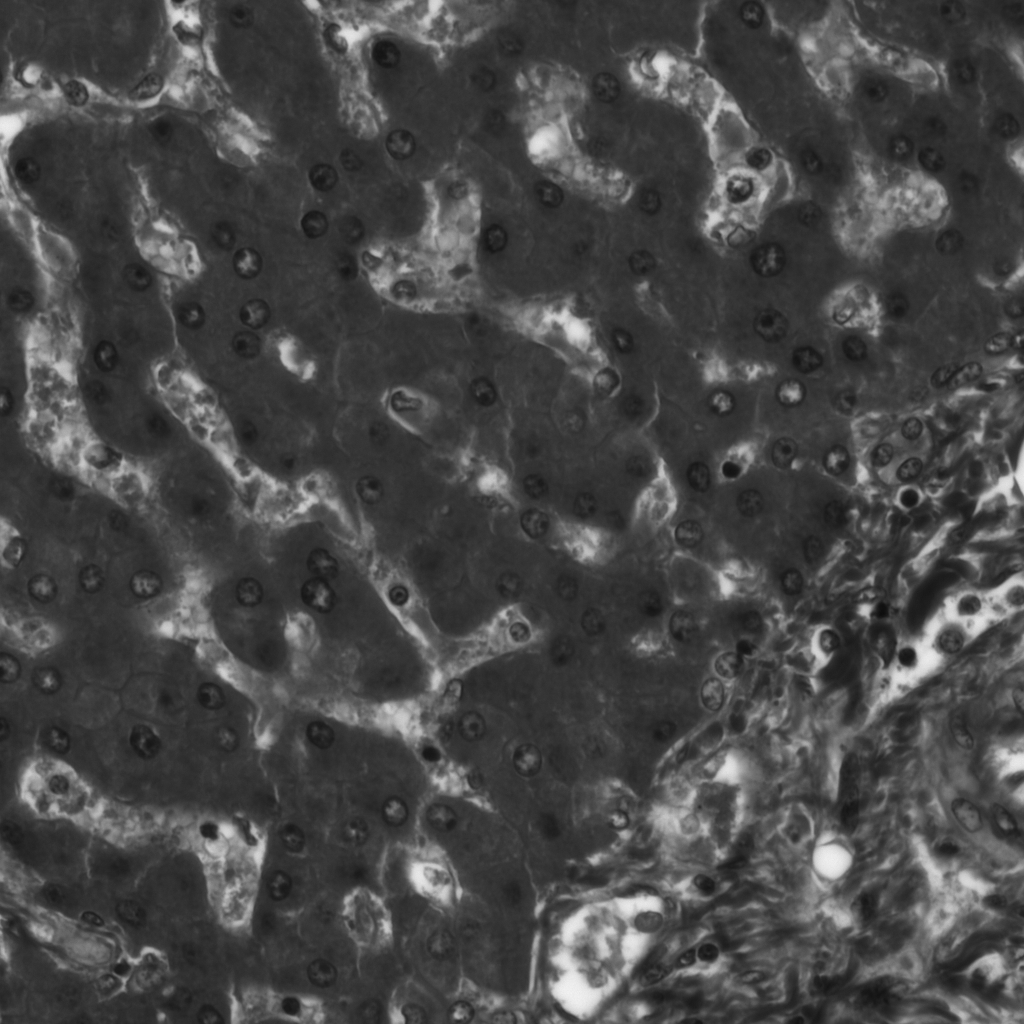}}  &
\subfigure[LR]{\includegraphics[width=0.84 in]{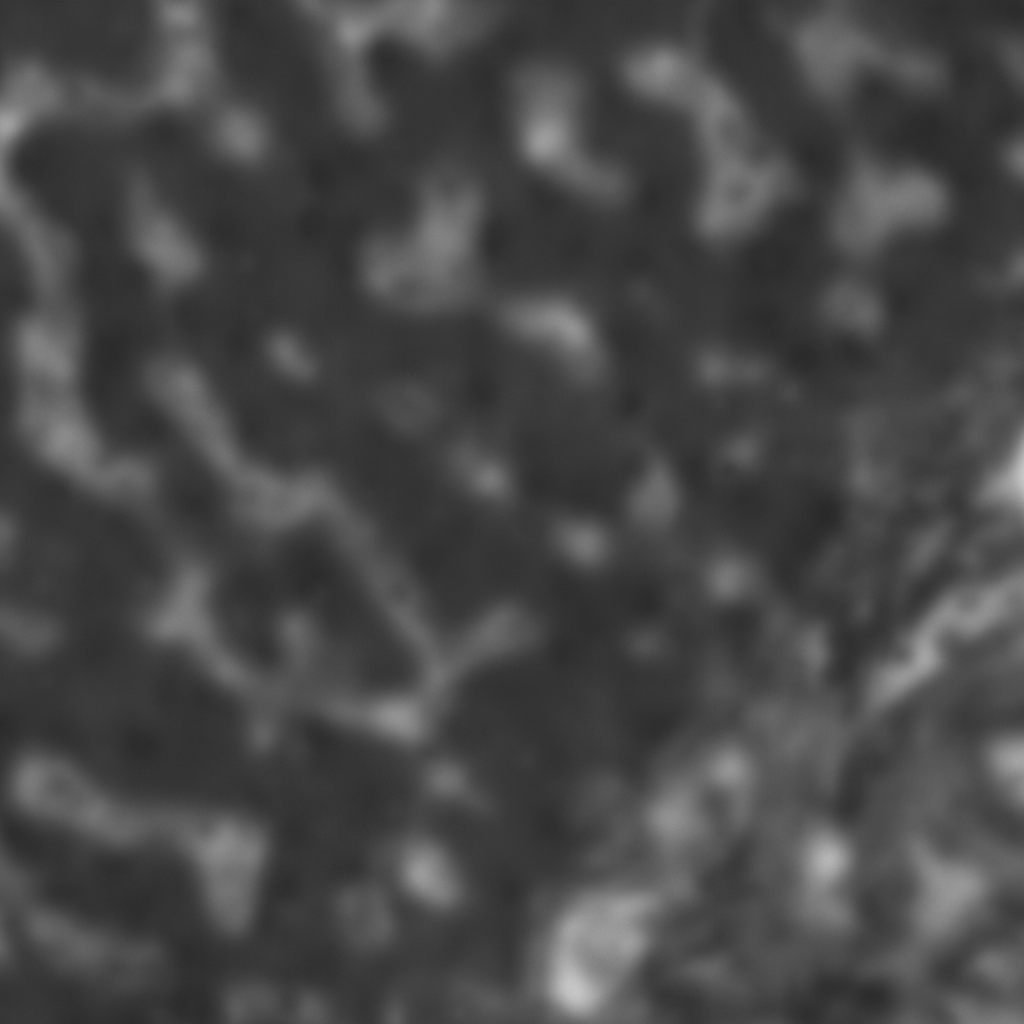}}   &
\subfigure[\textbf{GR ($k=7$)}]{\includegraphics[width=0.84 in]{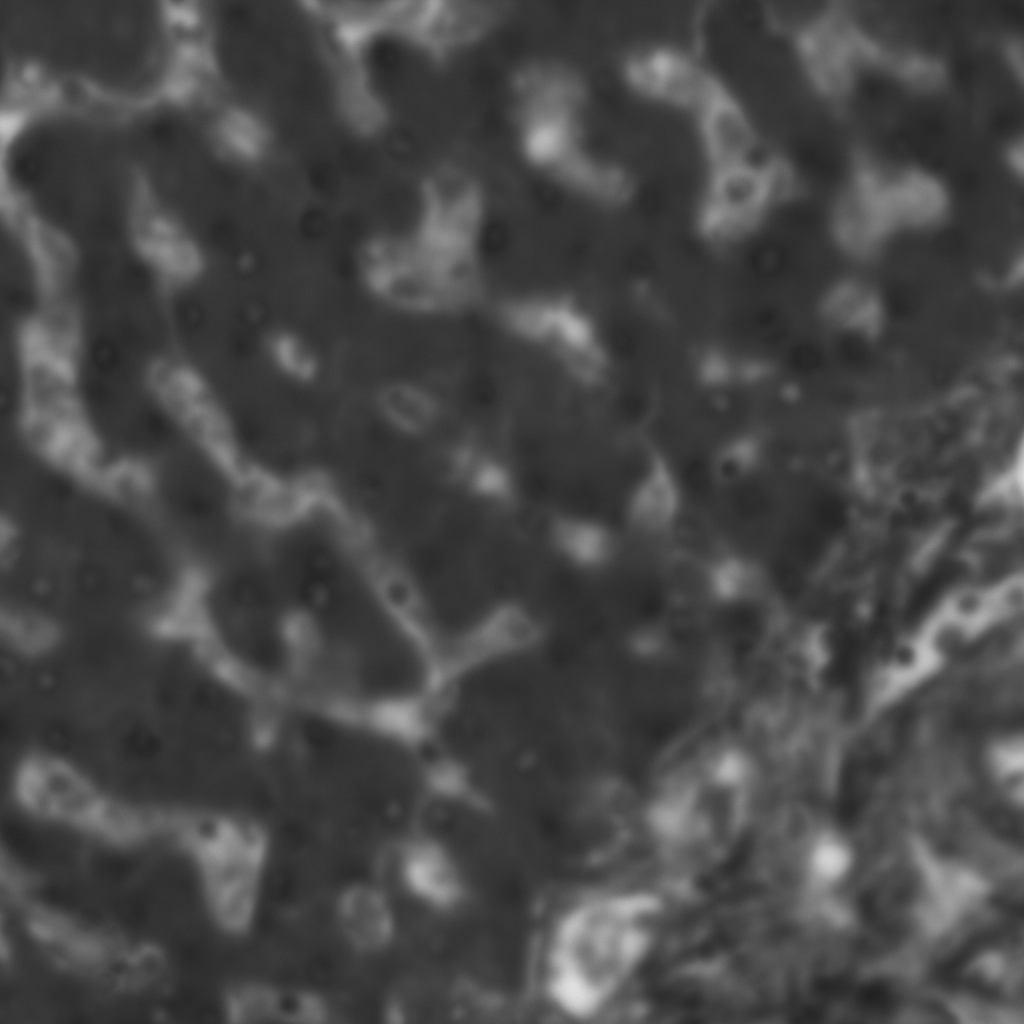}}&
\subfigure[GR ($k=9$)]{\includegraphics[width=0.84 in]{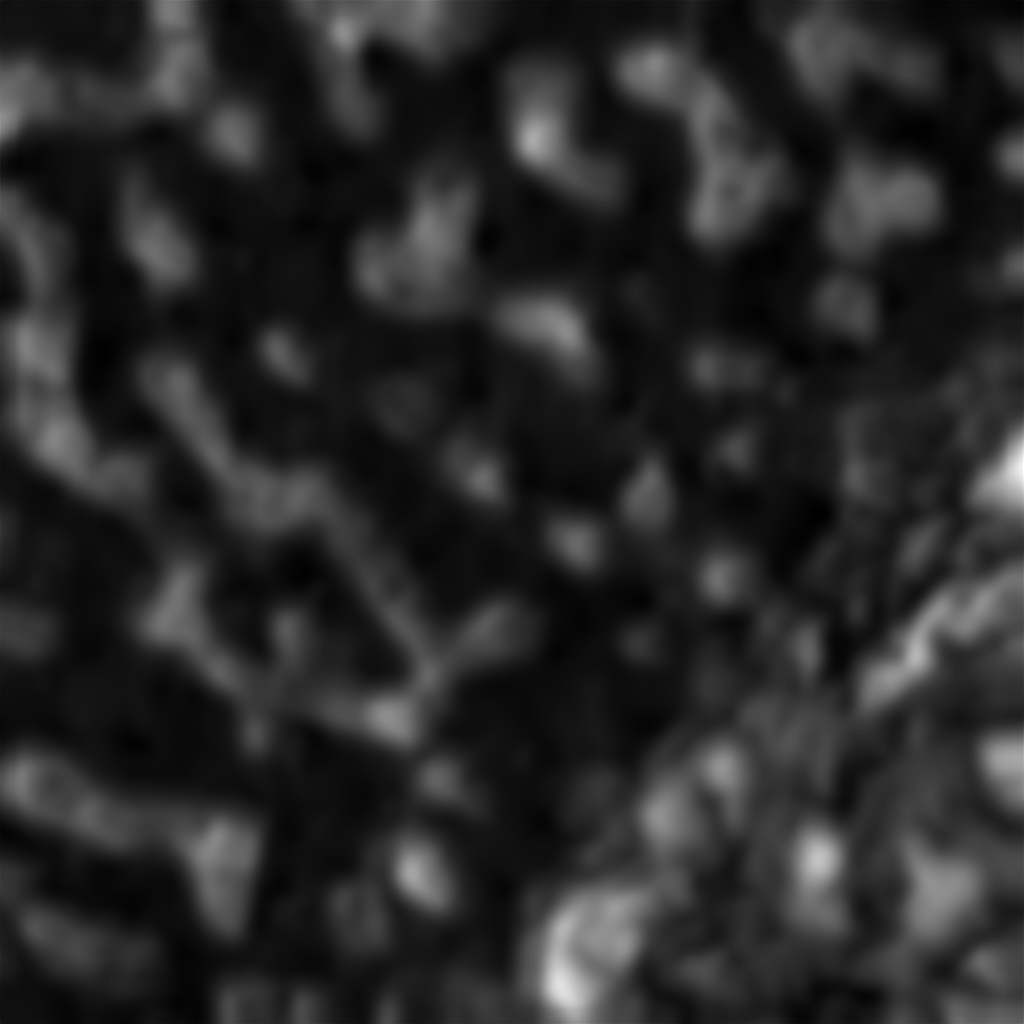}}   &
\subfigure[GR ($k=11$)]{\includegraphics[width=0.84 in]{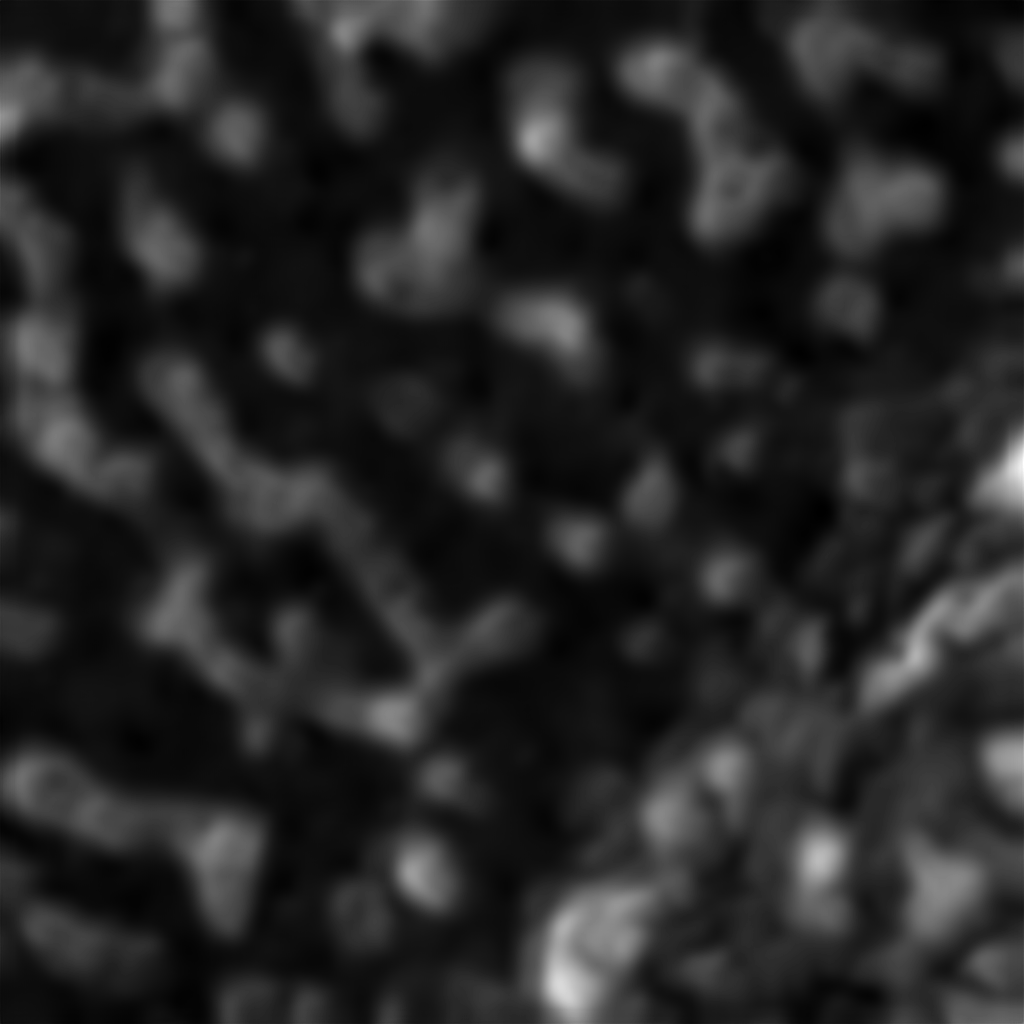}}   &
\subfigure[GR ($k=13$)]{\includegraphics[width=0.84 in]{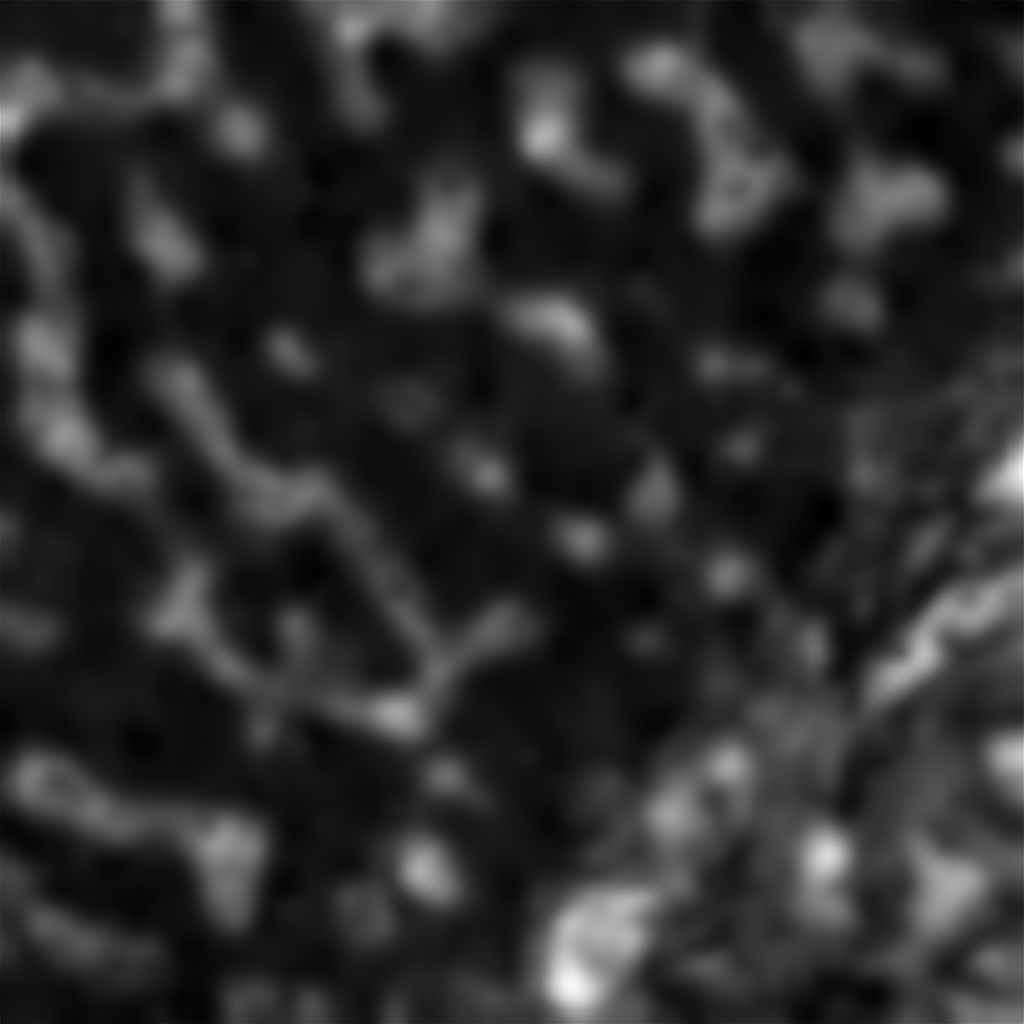}}   &
\subfigure[CDL]{\includegraphics[width=0.84 in]{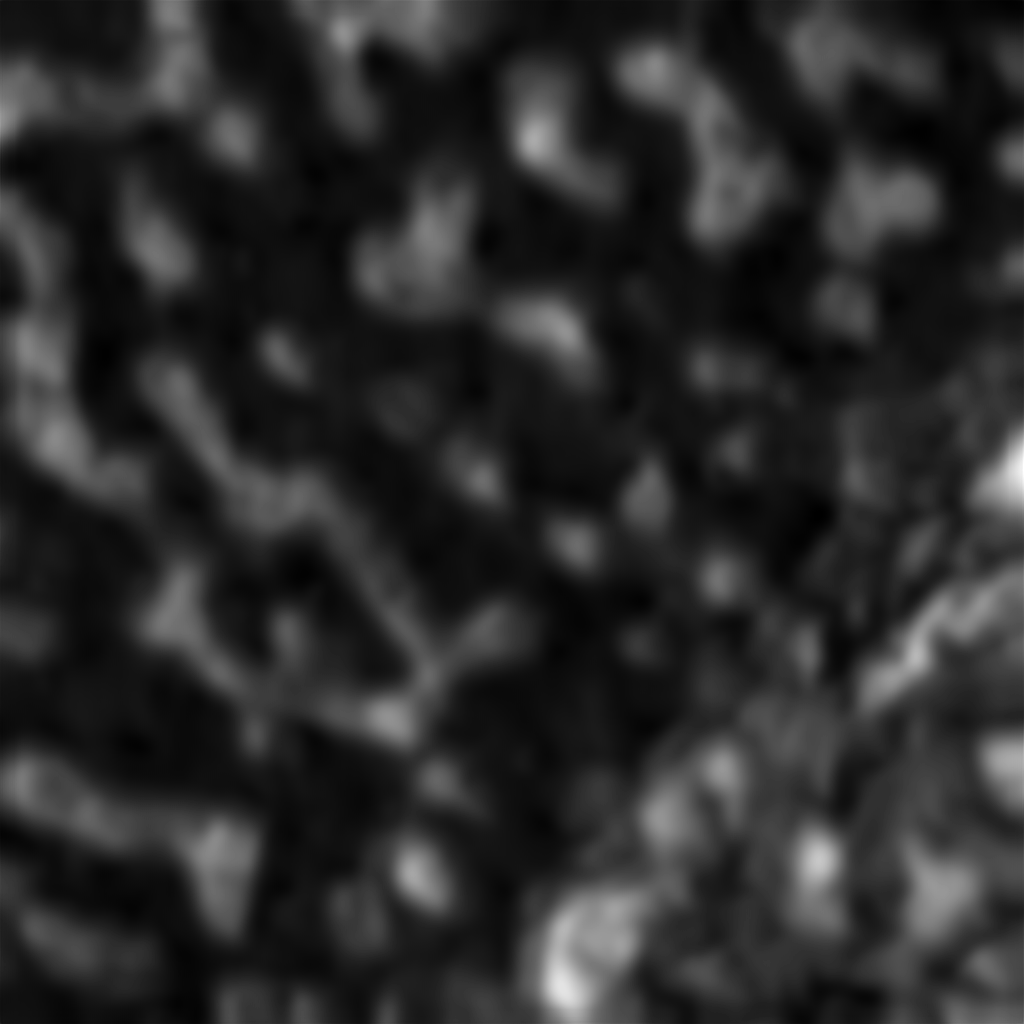}}  \\
\subfigure[HR]{\includegraphics[width=0.84 in]{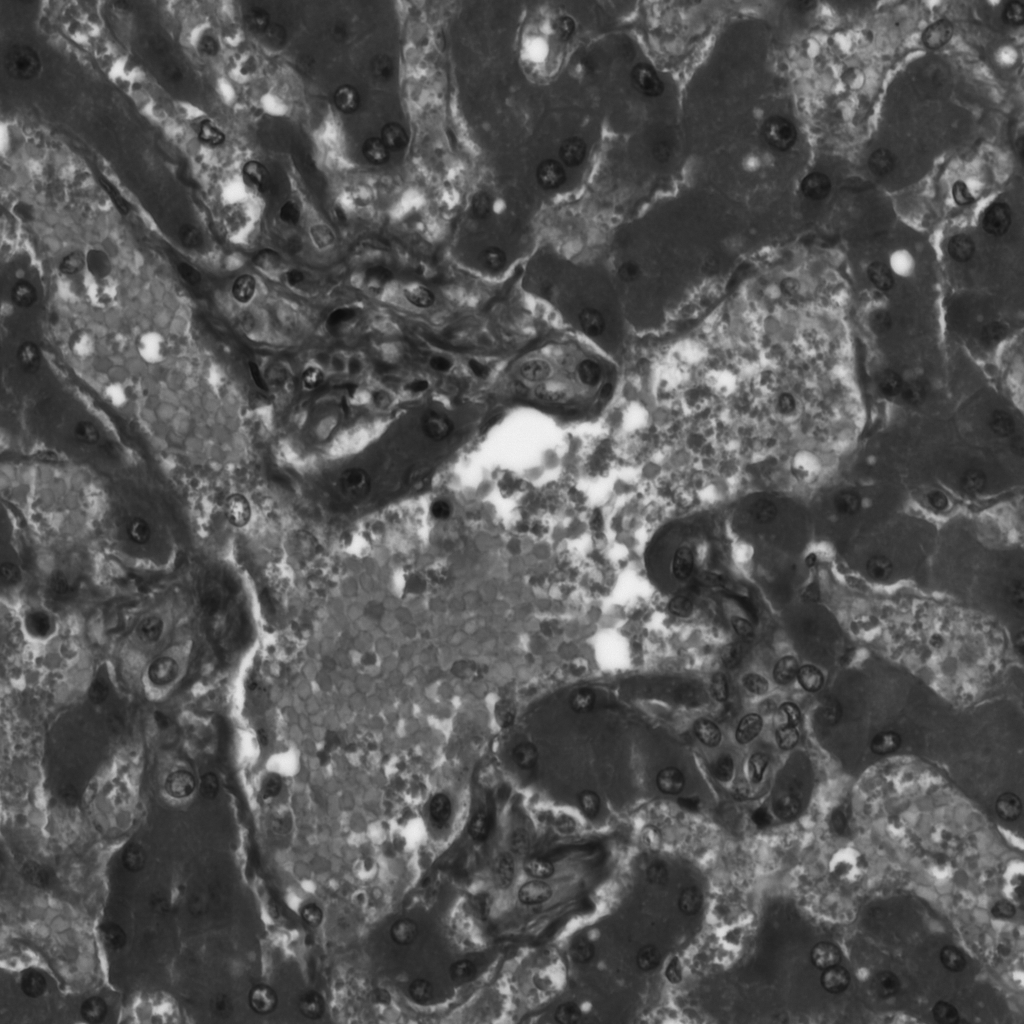}}  &
\subfigure[LR]{\includegraphics[width=0.84 in]{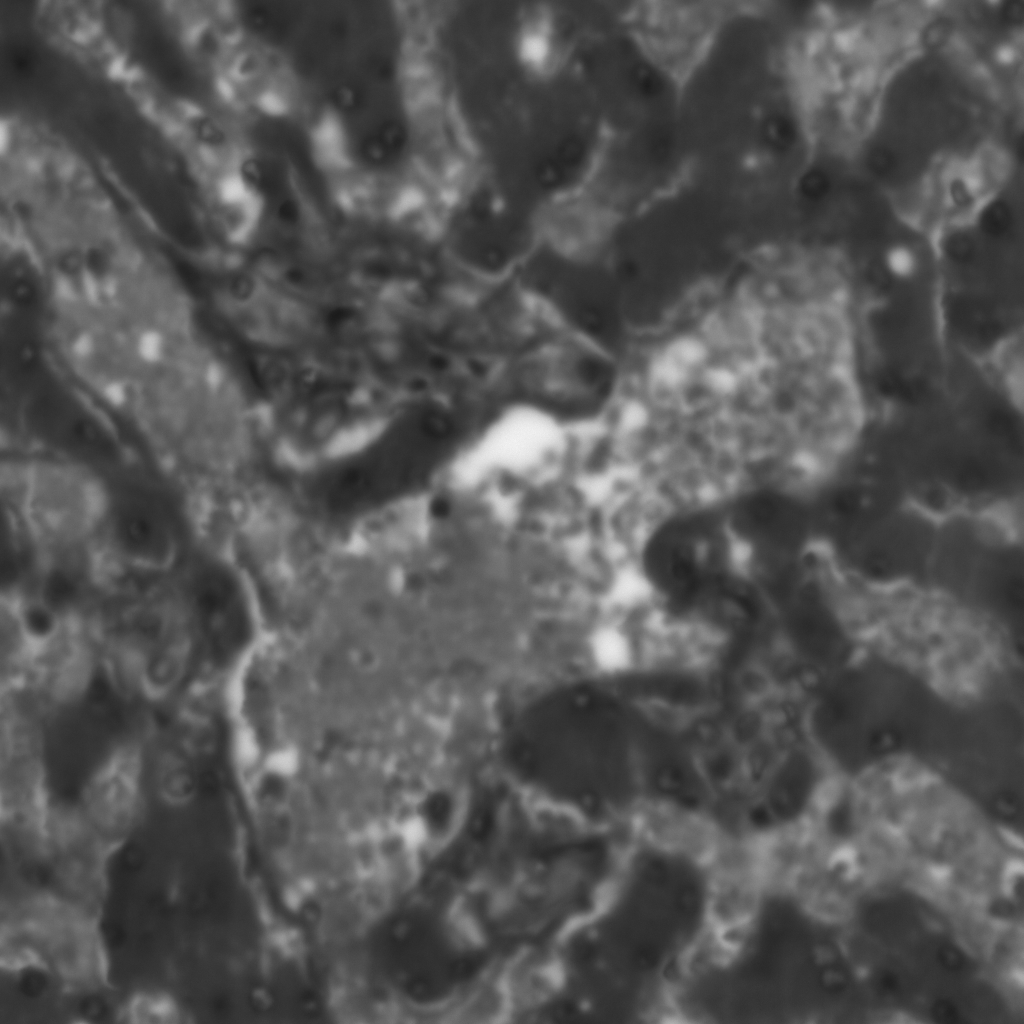}}   &
\subfigure[\textbf{GR ($k=7$)}]{\includegraphics[width=0.84 in]{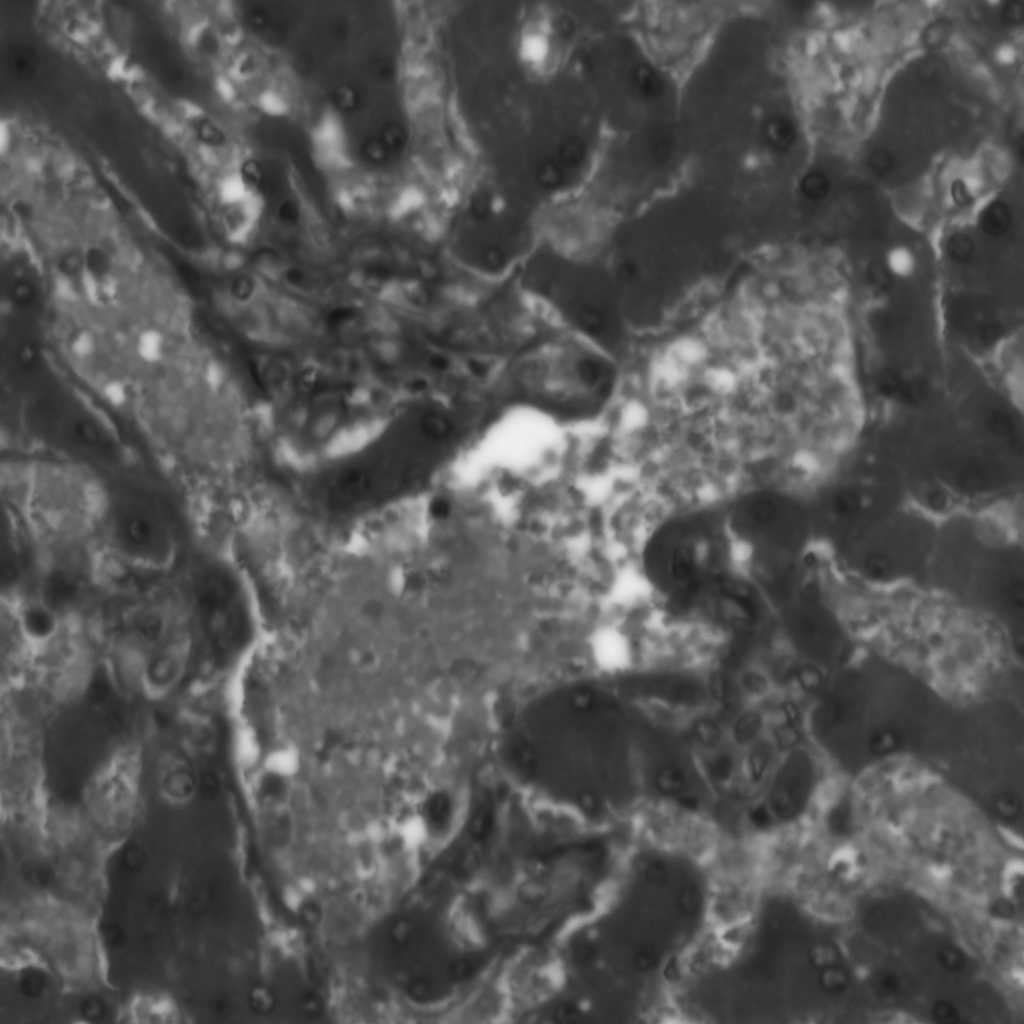}}&
\subfigure[GR ($k=9$)]{\includegraphics[width=0.84 in]{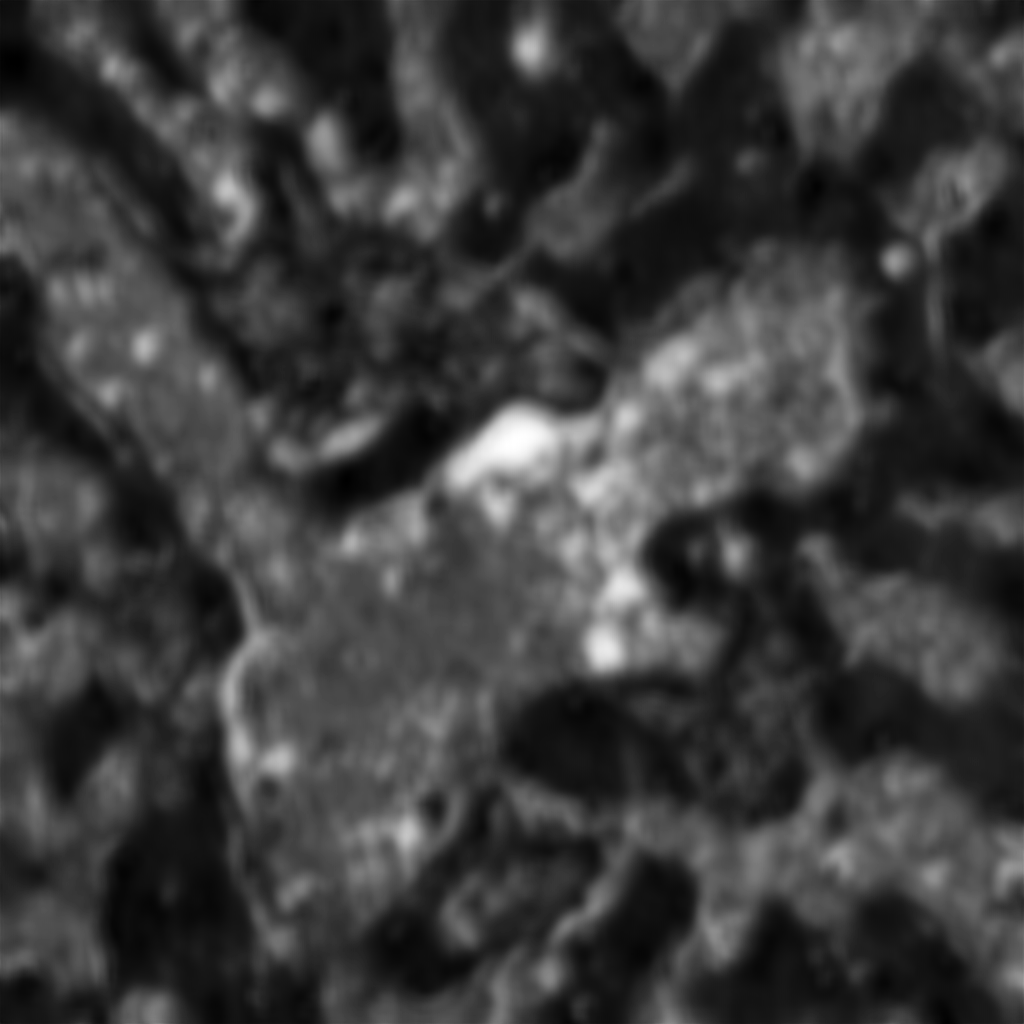}}   &
\subfigure[GR ($k=11$)]{\includegraphics[width=0.84 in]{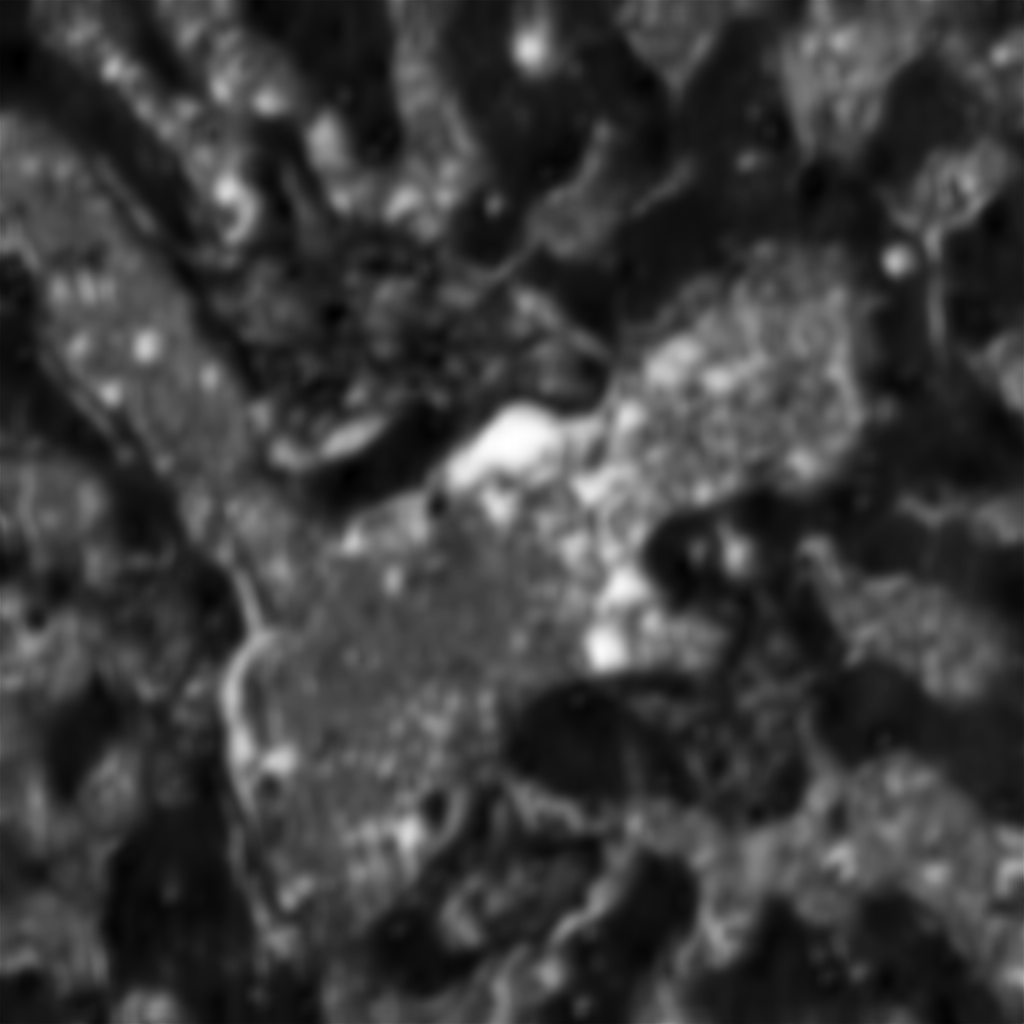}}   &
\subfigure[GR ($k=13$)]{\includegraphics[width=0.84 in]{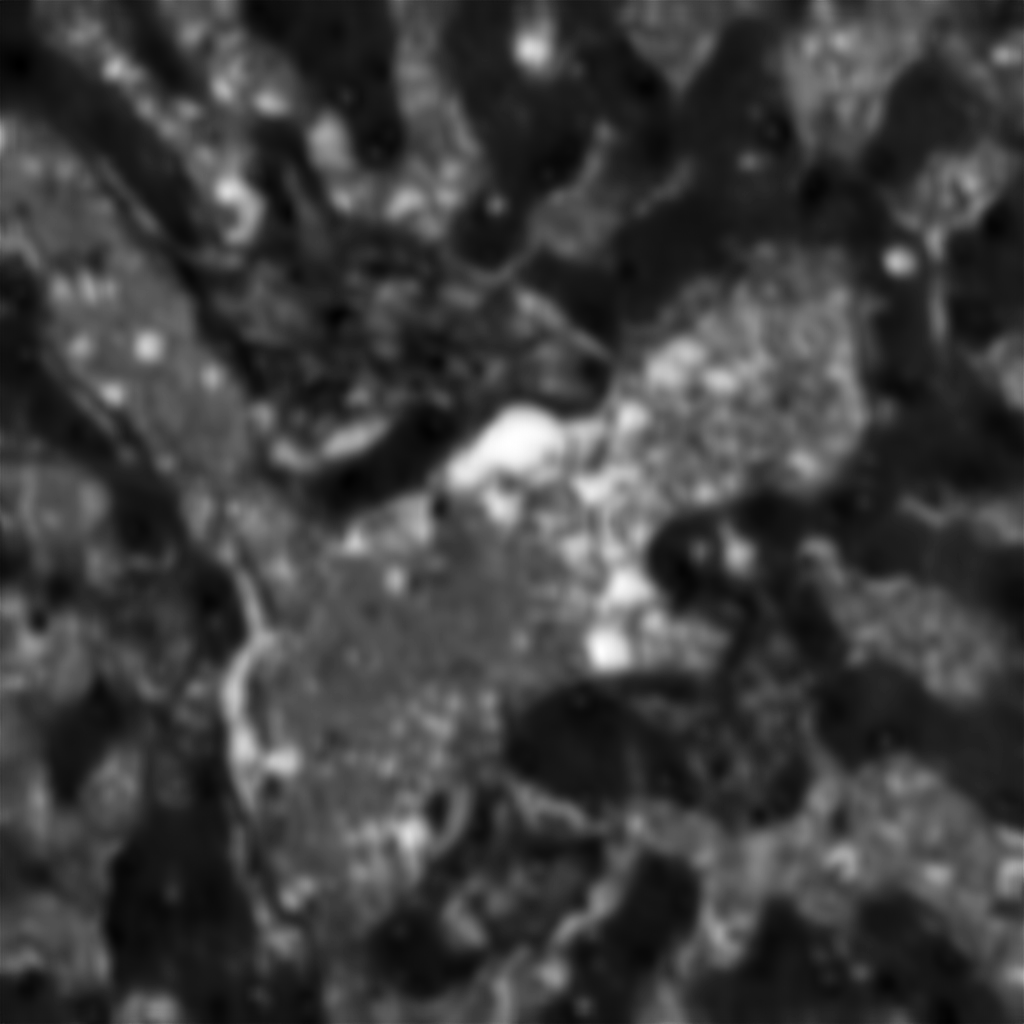}}   &
\subfigure[CDL]{\includegraphics[width=0.84 in]{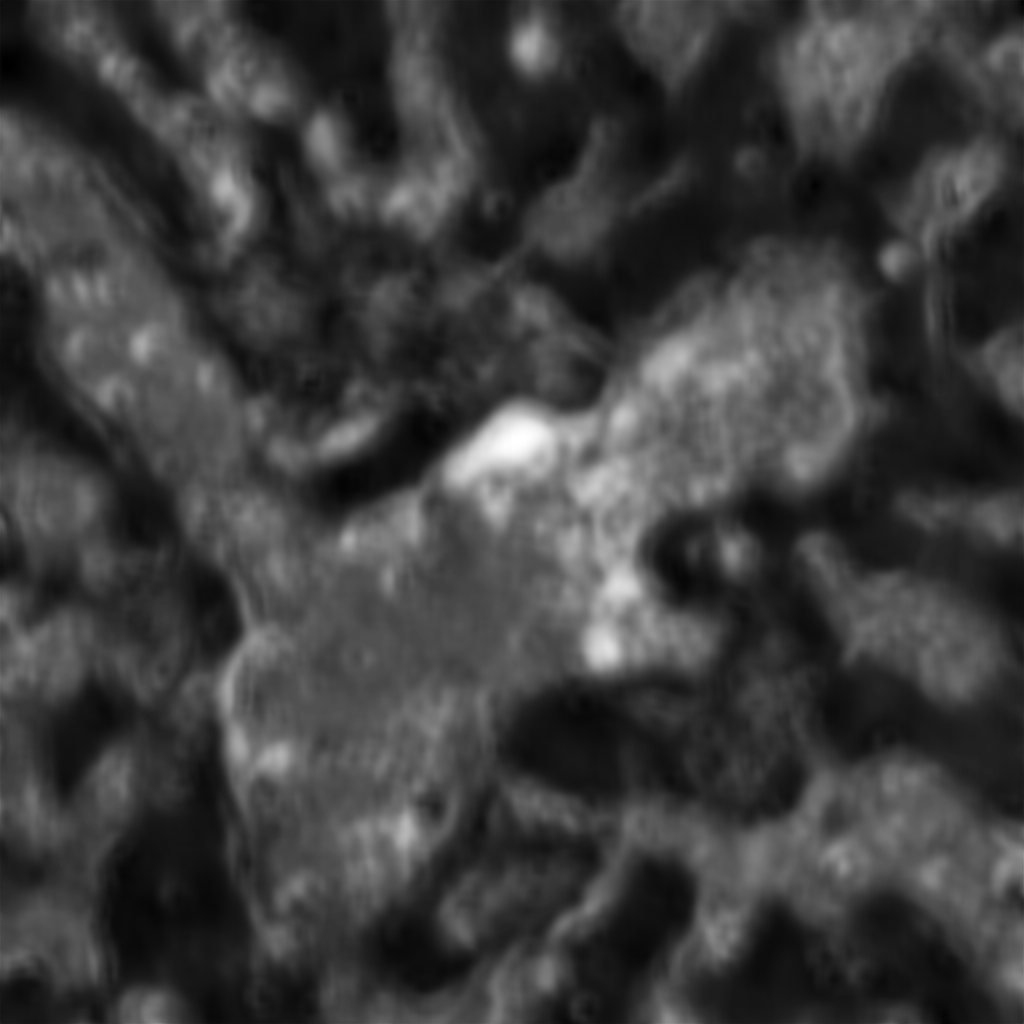}}  \\
\subfigure[HR]{\includegraphics[width=0.84 in]{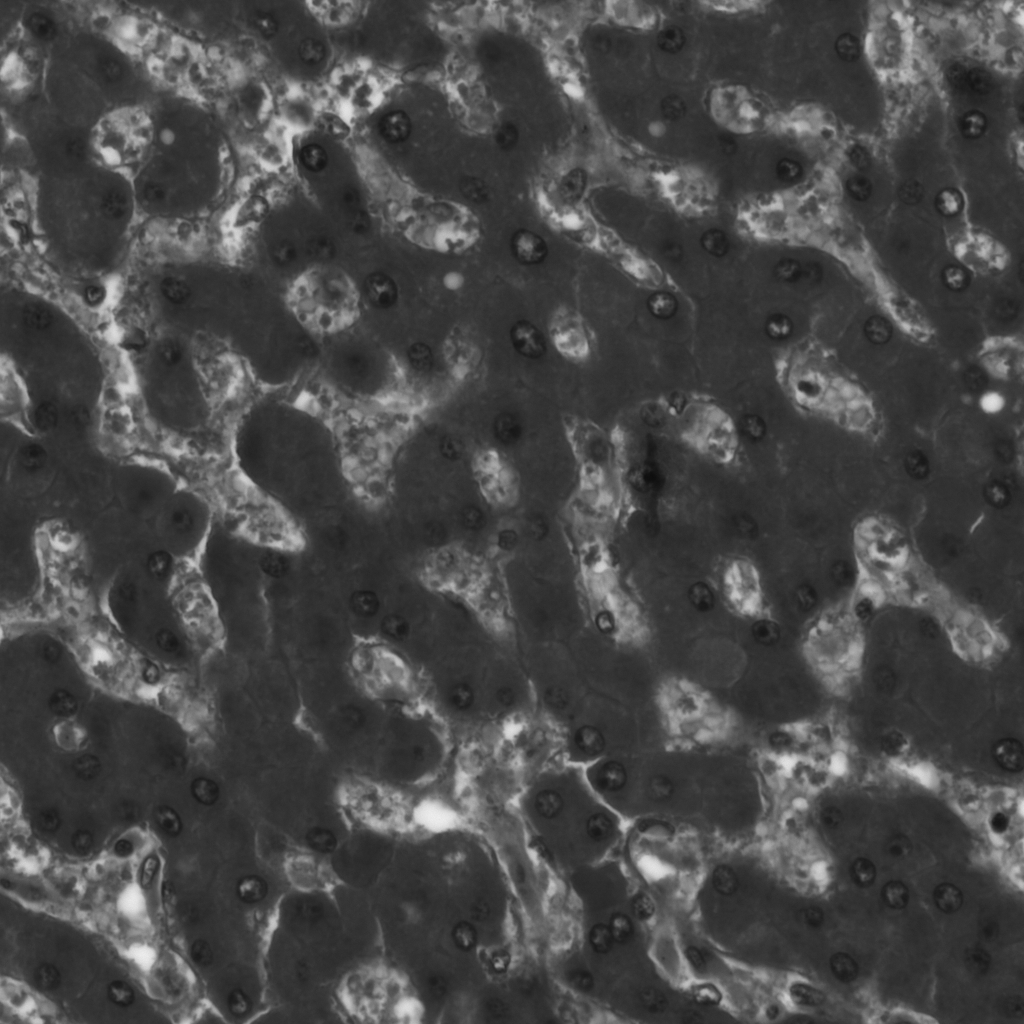}}  &
\subfigure[LR]{\includegraphics[width=0.84 in]{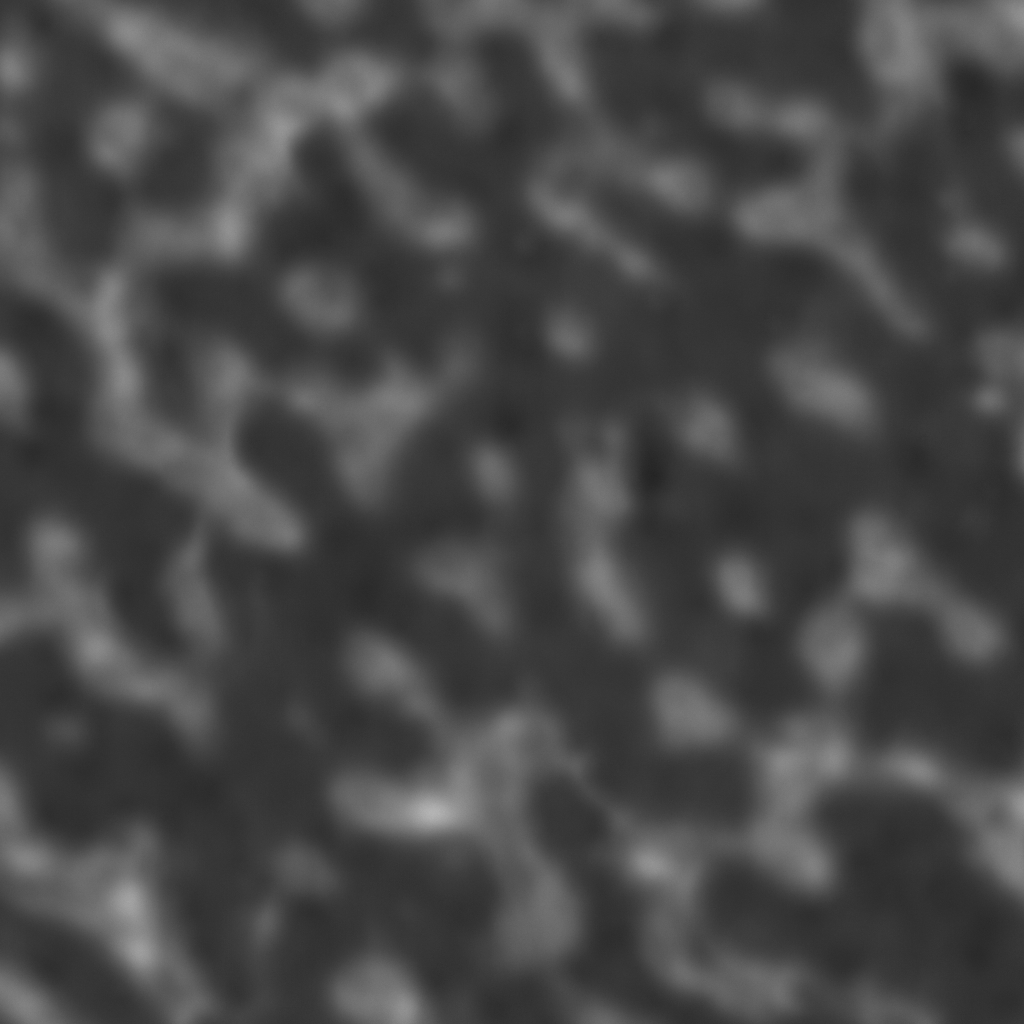}}   &
\subfigure[\textbf{GR ($k=7$)}]{\includegraphics[width=0.84 in]{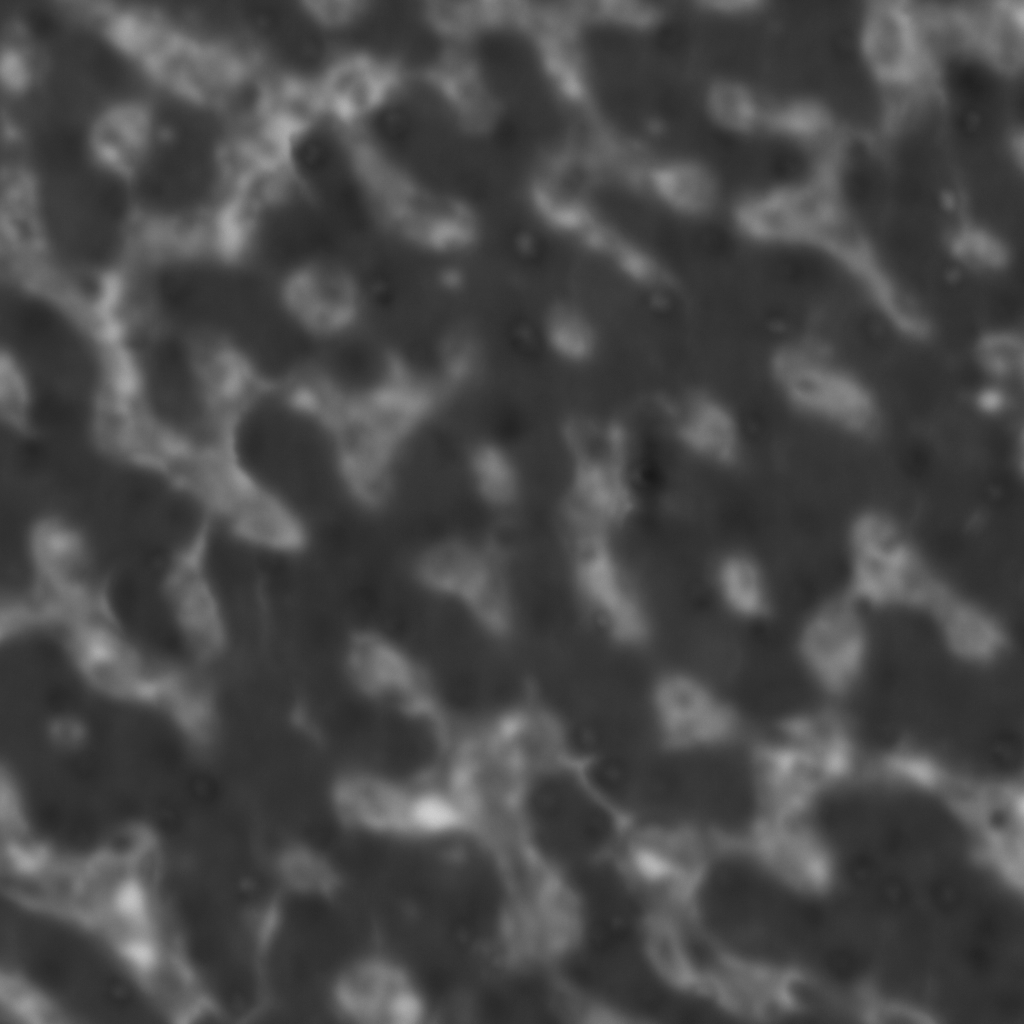}}&
\subfigure[GR ($k=9$)]{\includegraphics[width=0.84 in]{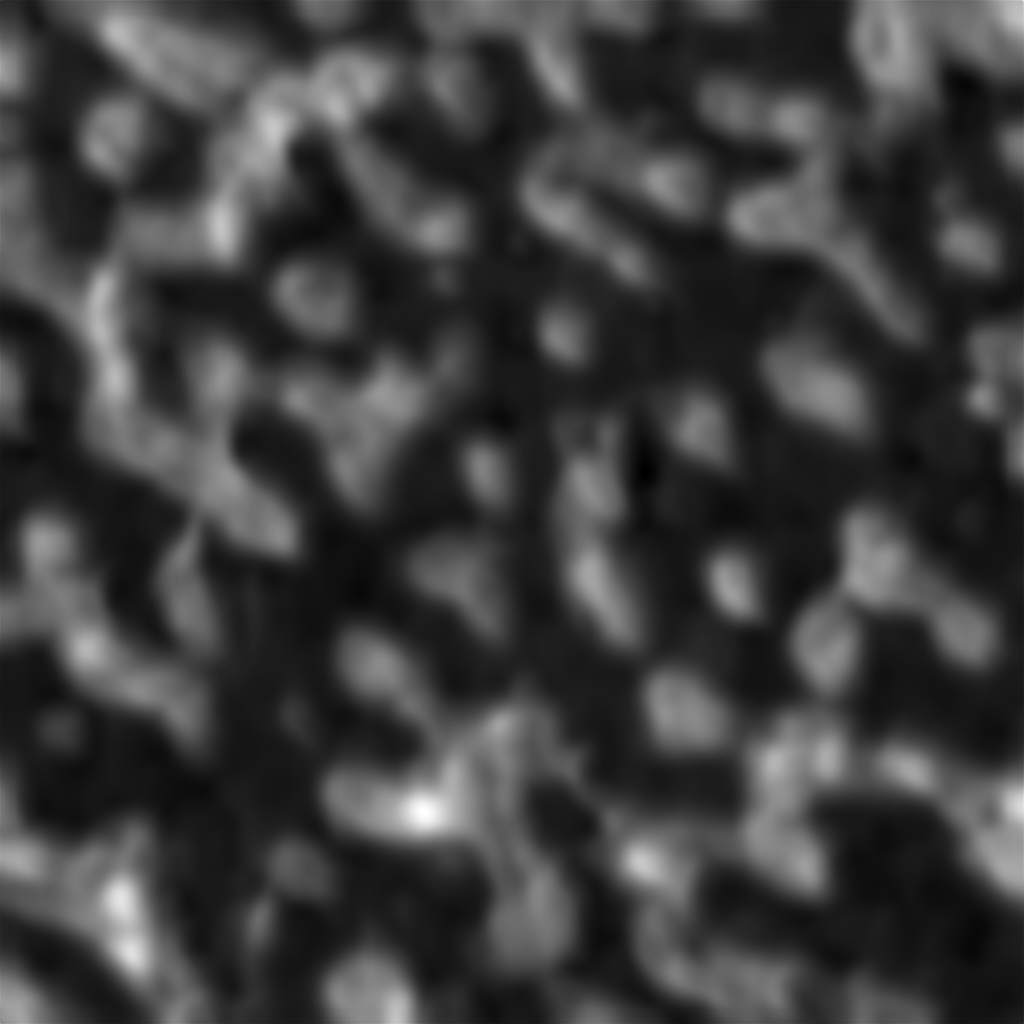}}   &
\subfigure[GR ($k=11$)]{\includegraphics[width=0.84 in]{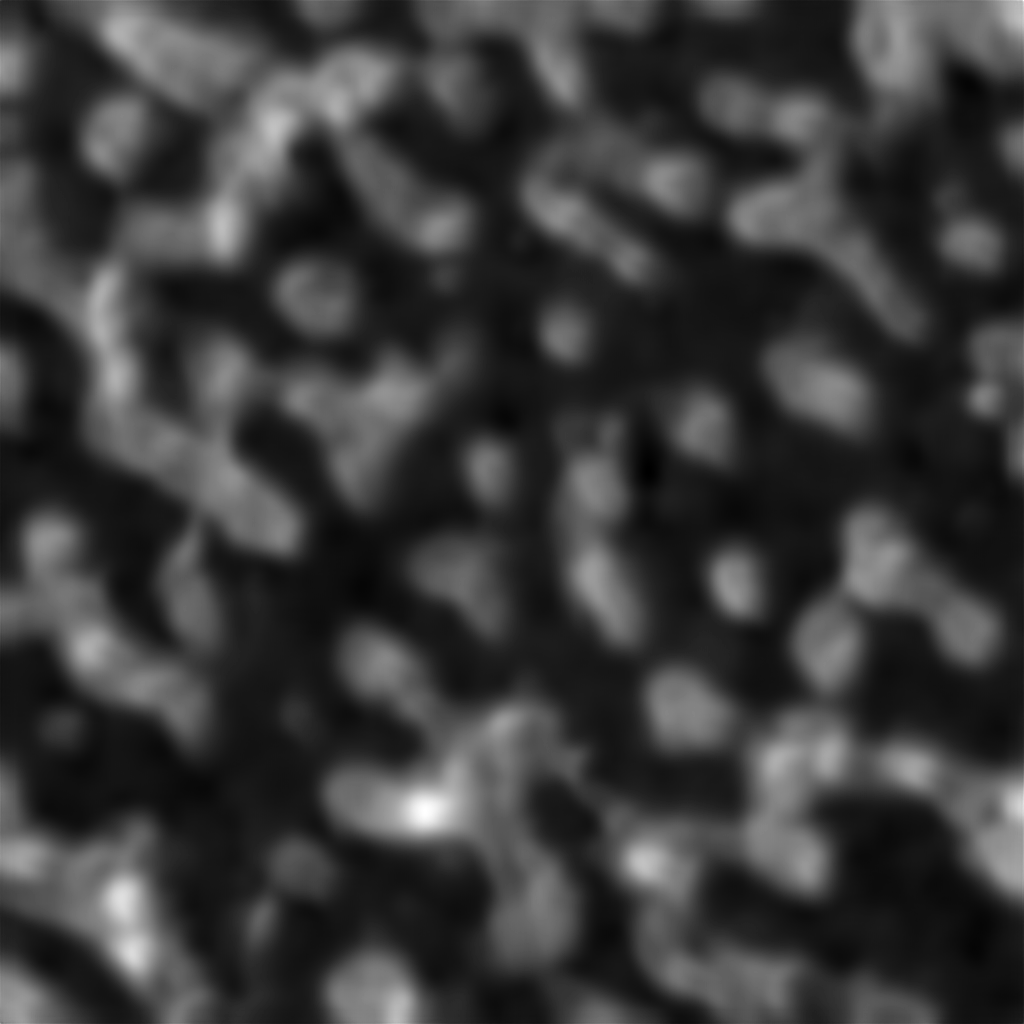}}   &
\subfigure[GR ($k=13$)]{\includegraphics[width=0.84 in]{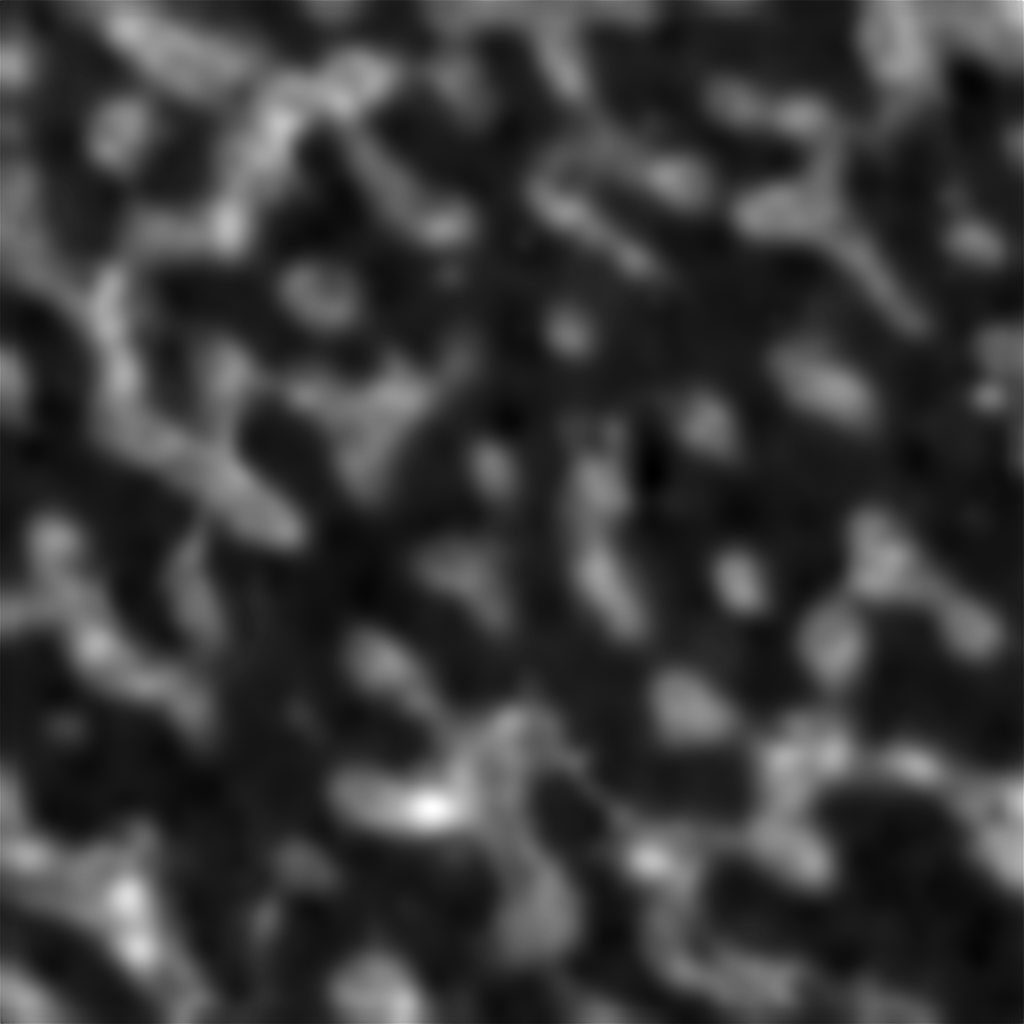}}   &
\subfigure[CDL]{\includegraphics[width=0.84 in]{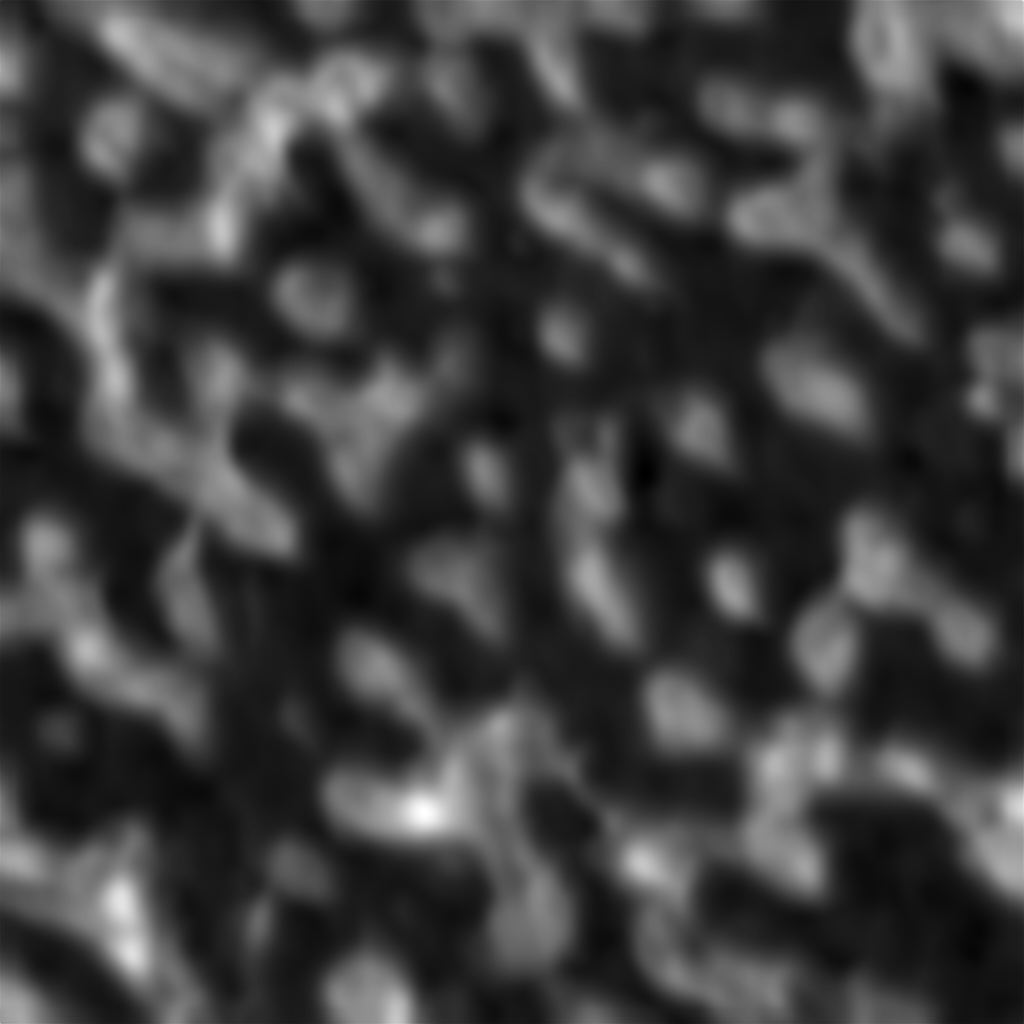}} 
\end{tabular}
\caption{Visual comparison of BME-GR with varying kernel sizes ($k$) on the FocusPath dataset. Results are shown for three different images: $S_1$ (first row), $S_2$ (second row), and $S_3$ (third row). Columns represent: (a, h, o) HR ground truth, (b, i, p) LR input, (c, j, q) BME-GR ($k=7$), (d, k, r) BME-GR ($k=9$), (e, l, s) BME-GR ($k=11$), (f, m, t) BME-GR ($k=13$), and (g, n, u) CDL results.}
\label{fig:Result_focuss_cross_val}
\end{center}
\end{figure*}

For training data construction, we establish image pairs by grouping specimens from identical slides at focal level $z_i$ as LR inputs, paired with corresponding HR targets from the same slide at focal plane $z = 8$. This pairing strategy yields approximately six training pairs per experimental group, with the remaining pairs used for testing. We train our BME-GR algorithm and benchmark it against the CDL algorithm using optimized hyperparameters. Table~\ref{tab:metrics_focusspath} presents quantitative results for three representative test images from Slide 1 at different strips and positions: (Strip 1, Position 3), defined as $S_1$; (Strip 0, Position 2), defined as $S_2$; and (Strip 0, Position 1), defined as $S_3$ (cf. Hosseine et al.~\cite{focusspath}). Figure~\ref{fig:Result_focuss_cross_val} illustrates performance metrics across varying blur with different $k$s. Notably, for all three test images, we obtain the best cross-validation results for $k = 7$, as shown qualitatively in Figure~\ref{fig:Result_focuss_cross_val} and quantitatively supported by Table~\ref{tab:metrics_focusspath}, demonstrating that our method produces better results than those obtained using the CDL algorithm.

\begin{table}[t!]
\centering
\caption{PSNR and SSIM values for different $k$ values}\small
\begin{tabular}{|c|c|c|c|c|c|c|}
\hline
Image & Metric & \multicolumn{4}{c|}{$k$ values} & CDL \\
\cline{3-6}
& & $k=13$ & $k=11$ & $k=9$ & $k=7$ &  \\
\hline
\multirow{2}{*}{$S_1$} & PSNR & 29.45 & 29.33 & 31.02 & \textbf{32.65} & 29.30 \\
& SSIM & 0.84 & 0.84 & 0.88 & \textbf{0.91} & 0.83 \\
\hline
\multirow{2}{*}{$S_2$} & PSNR & 26.84 & 26.12 & 25.93 & \textbf{27.93} & 26.20 \\
& SSIM & 0.83 & 0.82 & 0.80 & \textbf{0.86} & 0.81 \\
\hline
\multirow{2}{*}{$S_3$} & PSNR & 20.12 & 21.22 & 22.81 & \textbf{22.92} & 21.10 \\
& SSIM & 0.55 & 0.54 & 0.54 & \textbf{0.55} & 0.53 \\
\hline
\end{tabular}
\label{tab:metrics_focusspath}
\end{table}


\section{Deblurring via Dictionary Learning for No-Correspondence Data}
\label{sec:NC_data}
This section extends the blur-coupled dictionary learning framework to address the challenging scenario where no correspondence exists between LR and HR images. While traditional methods require aligned image pairs, real-world applications often lack such correspondence due to practical constraints in data acquisition. Our approach leverages the underlying relationships between HR and LR domains under a linear blur degradation model, where LR images are generated through convolution with a fixed blur kernel $\mathbf{b}$.

Consider the dataset in \eqref{eq:unpaired_data}
with no pairwise relationships between HR patches $\mathbf{X}^{h} \in \mathbb{R}^{N_{h} \times N}$ and LR patches $\mathbf{Y}^{l} \in \mathbb{R}^{N_{l} \times M}$. The degradation process follows:
\begin{align}
\mathbf{Y}^{l} = \mathbf{B}\mathbf{D}^{h}\mathbf{C}, \quad \text{and} \quad 
\mathbf{X}^{h} = \mathbf{D}^{h}\widetilde{\mathbf{C}}, \notag
\end{align}
where $\mathbf{C}$, $\widetilde{\mathbf{C}}$ are matrices with sparse codes. This formulation acknowledges that while both domains share the same underlying dictionary $\mathbf{D}^{h}$, their sparse representations differ due to the absence of explicit patch-level correspondences. This formulation addresses two key theoretical challenges:
\begin{enumerate}
    \item \textit{Shared Dictionary Consistency}: The common HR dictionary $\mathbf{D}^{h}$ enables cross-domain feature alignment by capturing fundamental image structures invariant to degradation. This leverages sparse representation principles where dictionaries encode cross-domain features.
    \item \textit{Structural Regularisation}: The block-Toeplitz structure of $\mathbf{B}$ provides inherent mathematical regularisation, constraining solutions to physically feasible blur operators despite missing correspondences.
\end{enumerate}

\subsection{Problem Formulation For No Correspondence Data}
For the no-correspondence case, learning $\mathbf{B, \ D}^{h}$, solves the optimization problem,
\begin{align}
\min_{\mathbf{B}, \mathbf{D}^{h}, \mathbf{C}, \widetilde{\mathbf{C}}} & \  \left\| \mathbf{Y}^{l} - \mathbf{BD}^{h}\mathbf{C} \right\|_{F}^{2} + \left\| \mathbf{X}^{h} - \mathbf{D}^{h}\widetilde{\mathbf{C}} \right\|_{F}^{2} \nonumber \\
& + \lambda \left( \| \mathbf{C} \|_{1,1} + \| \widetilde{\mathbf{C}} \|_{1,1} \right).
\label{eq:NonCorespondence-Opt-Poblem}
\end{align}
The problem is inherently non-convex, which makes it difficult to solve directly in order to address the non-correspondence between HR and LR images. However, it can be divided into smaller, well-structured convex optimization sub-problems that may be addressed iteratively in an alternating method. Specifically, the solution alternates between three convex subproblems:

\subsubsection{Sparse code recovery}
For fixed $\mathbf{D}^{h}$ and $\mathbf{B}$, the sparse codes $\mathbf{C}$ and $\widetilde{\mathbf{C}}$ are recovered by solving the following $\ell_1$ minimization problems:
\begin{align}
&\min_{\mathbf{C}} \;
\left\|\mathbf{Y}^{l} - \mathbf{B}\mathbf{D}^{h}\mathbf{C}\right\|_{F}^{2} 
+ \lambda \|\mathbf{C}\|_{1,1},\notag \\
&\min_{\widetilde{\mathbf{C}}}\left\|\mathbf{X}^{h} - \mathbf{D}^{h}\widetilde{\mathbf{C}}\right\|_{F}^{2} 
+ \lambda \|\widetilde{\mathbf{C}}\|_{1,1}.
\label{eq:NC_FISTA}
\end{align}

\subsubsection{Blur matrix estimation}
$\mathbf{B}$ is estimated by considering the low-resolution fidelity term
\begin{align}
    \min_{\mathbf{B}} \,  \left\| \mathbf{Y}^{l} - \mathbf{BD}^{h}\mathbf{C} \right\|_{F}^{2}. \label{eq:NC_B_estimate}
\end{align}
for fixed $\mathbf{D}^{h}$ and $\mathbf{C}$. This formulation is identical to the blur matrix estimation problem introduced in Section~\ref{sec:Problemformulation} for the correspondence case and it can be solved using either BME-GR or BME-SR. Since BME-SR consistently provides better performance and as for the non-correspondence setting, the estimation of $\mathbf{B}$ does not require HR patches, we adopt BME-SR for blur matrix estimation in this non-correspondence setting as well.

\begin{algorithm}[!t]
\caption{Joint Dictionary Learning via K-SVD}
\label{alg:joint-dict-update}
    \begin{algorithmic}[1]
        \Require $\mathbf{Y}^l$, $\mathbf{X}^h$, $\mathbf{B}$ \\
            \textbf{Intialize:} $\mathbf{D}^h$, $\mathbf{C}$, $\widetilde{\mathbf{C}}$
        \Ensure Updated dictionary $\mathbf{D}^h$ and sparse codes $\mathbf{C}$, $\widetilde{\mathbf{C}}$
        \For{$t = 1, 2, \ldots , N_c $}
            \State \textbf{Phase I: Extract Support and Coefficients}
            \State $\omega_t \leftarrow \{i : \mathbf{C}_{t,i} \neq 0\}$ \Comment{Support of $t$-th row in $\mathbf{C}$}
            \State $\widetilde{\omega}_t \leftarrow \{j : \widetilde{\mathbf{C}}_{t,j} \neq 0\}$ \Comment{Support of $t$-th row in $\widetilde{\mathbf{C}}$}
            
            \State \textbf{Phase II: Residual Computation}
            \State $\mathbf{D}^h_{:,t} \leftarrow \mathbf{0}$ \Comment{Remove $t$-th atom contribution}
            \State $\mathbf{E}^h_t \leftarrow \mathbf{X}^h_{:,\widetilde{\omega}_t} - \mathbf{D}^h \widetilde{\mathbf{C}}_{:,\widetilde{\omega}_t}$
            \State $\mathbf{E}^l_t \leftarrow (\mathbf{B}^T\mathbf{B})^{-1}\mathbf{B}^T(\mathbf{Y}^l_{:,\omega_t} - \mathbf{B}\mathbf{D}^h\mathbf{C}_{:,\omega_t})$
            
            \State \textbf{Phase III: Joint Optimization Setup} \Comment{Concatenate}
            \State $\mathbf{E}_t \leftarrow [\mathbf{E}^h_t \mid \mathbf{E}^l_t]$
            \State $\boldsymbol{\gamma}_t \leftarrow [\widetilde{\mathbf{C}}_{t,\widetilde{\omega}_t} \mid \mathbf{C}_{t,\omega_t}]$
            
            \State \textbf{Phase IV: SVD-Based Dictionary Update}
            \State $[\mathbf{U}, \boldsymbol{\Sigma}, \mathbf{V}^T] \leftarrow \text{SVD}(\mathbf{E}_t)$
            \State $\mathbf{D}^h_{:,t} \leftarrow \mathbf{U}_{:,1}$ \Comment{Update $t$-th dictionary atom}
            \State $\boldsymbol{\gamma}_t \leftarrow \boldsymbol{\Sigma}_{1,1} \mathbf{V}^T_{1,:}$ \Comment{Update coefficient vector}
            
            \State \textbf{Phase V: Sparse Code Update}
            \State $\widetilde{\mathbf{C}}_{t,\widetilde{\omega}_t} \leftarrow \boldsymbol{\gamma}_t[1:|\widetilde{\omega}_t|]$
            \State $\mathbf{C}_{t,\omega_t} \leftarrow \boldsymbol{\gamma}_t[(|\widetilde{\omega}_t|+1):\text{end}]$
        \EndFor
        \State \Return $\mathbf{D}^h$, $\mathbf{C}$, $\widetilde{\mathbf{C}}$
    \end{algorithmic}
\end{algorithm}

\subsubsection{Dictionary update}
In the non-correspondence framework, the $\mathbf{D}^h$ update step considers both the high- and low-resolution fidelity terms
\begin{align}
\min_{\mathbf{D}^{h}} \  \left\| \mathbf{Y}^{l} - \mathbf{BD}^{h}\mathbf{C} \right\|_{F}^{2} + \left\| \mathbf{X}^{h} -\mathbf{D}^{h}\widetilde{\mathbf{C}} \right\|_{F}^{2}.
\label{eq:NC-dictionary}
\end{align}
This is in contrast to the dictionary update step in the previous section for paired data, where only the high-resolution term is used. This is because in the correspondence case, the sparse codes are common to both the terms, whereas here they are not. The coupling between the two terms is the $\mathbf{D}^h$ term.

\begin{figure*}[!t]
\begin{center}
\begin{tabular}{cccc}
\subfigure[HR]{\includegraphics[width=1.45 in]{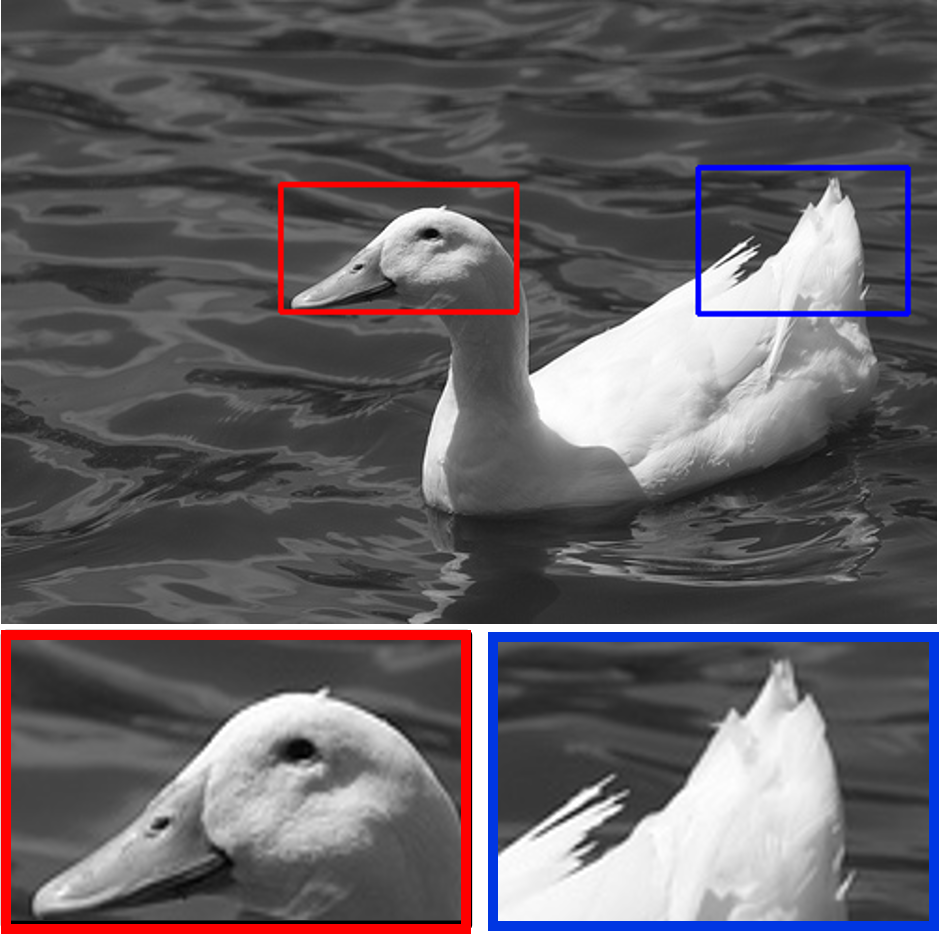}}   &
\subfigure[LR]{\includegraphics[width=1.45 in]{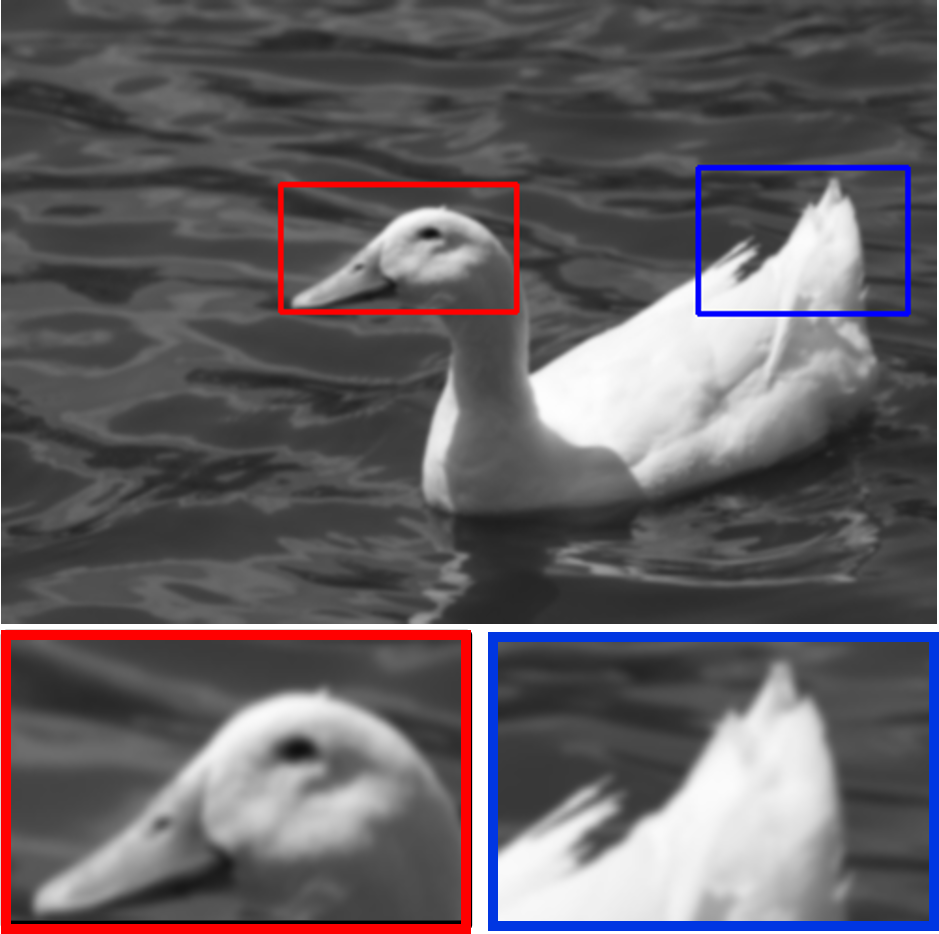}}    &
\subfigure[BME-SR (Scenario-1)]{\includegraphics[width=1.45 in]{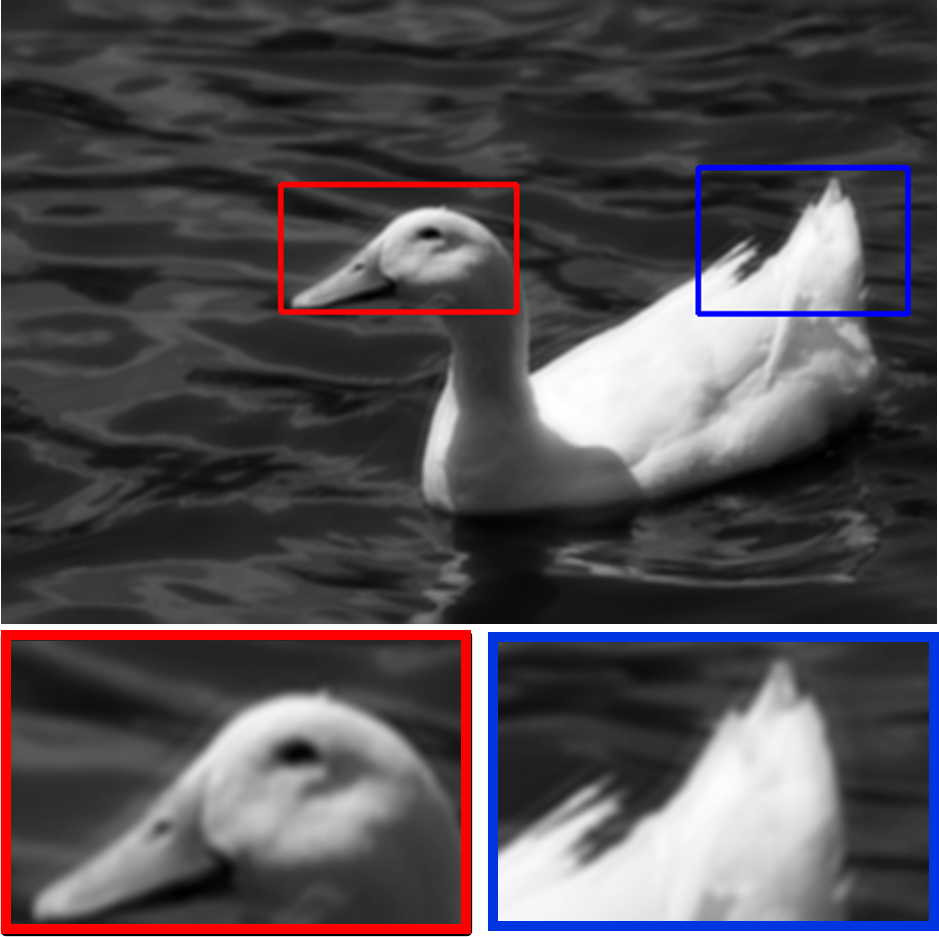}}    &
\subfigure[BME-SR (Scenario-2)]{\includegraphics[width=1.45 in]{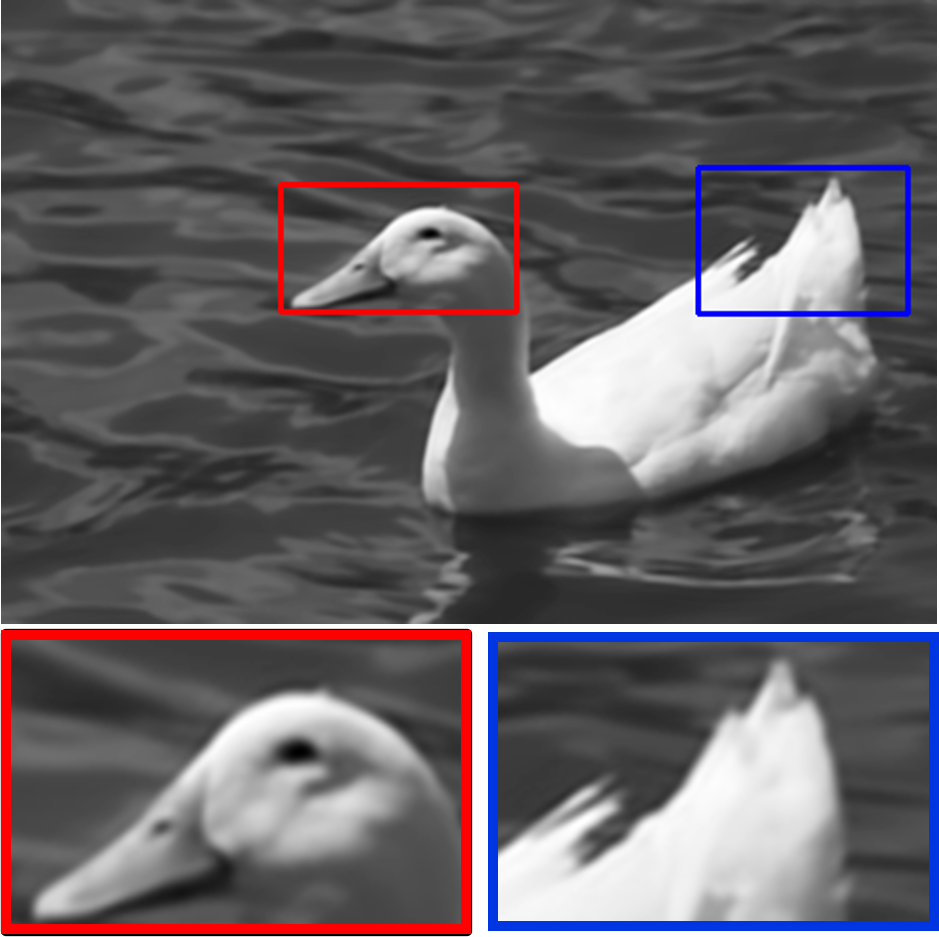}}    \\
\subfigure[HR]{\includegraphics[width=1.45 in]{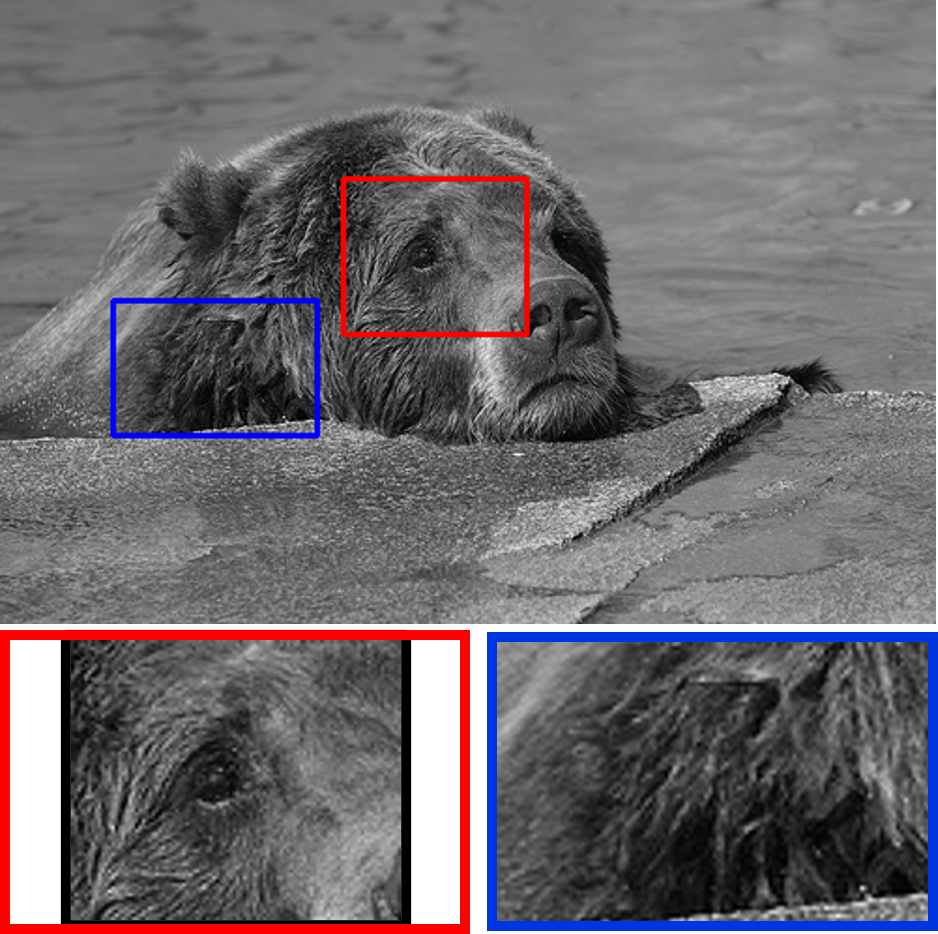}}   &
\subfigure[LR]{\includegraphics[width=1.45 in]{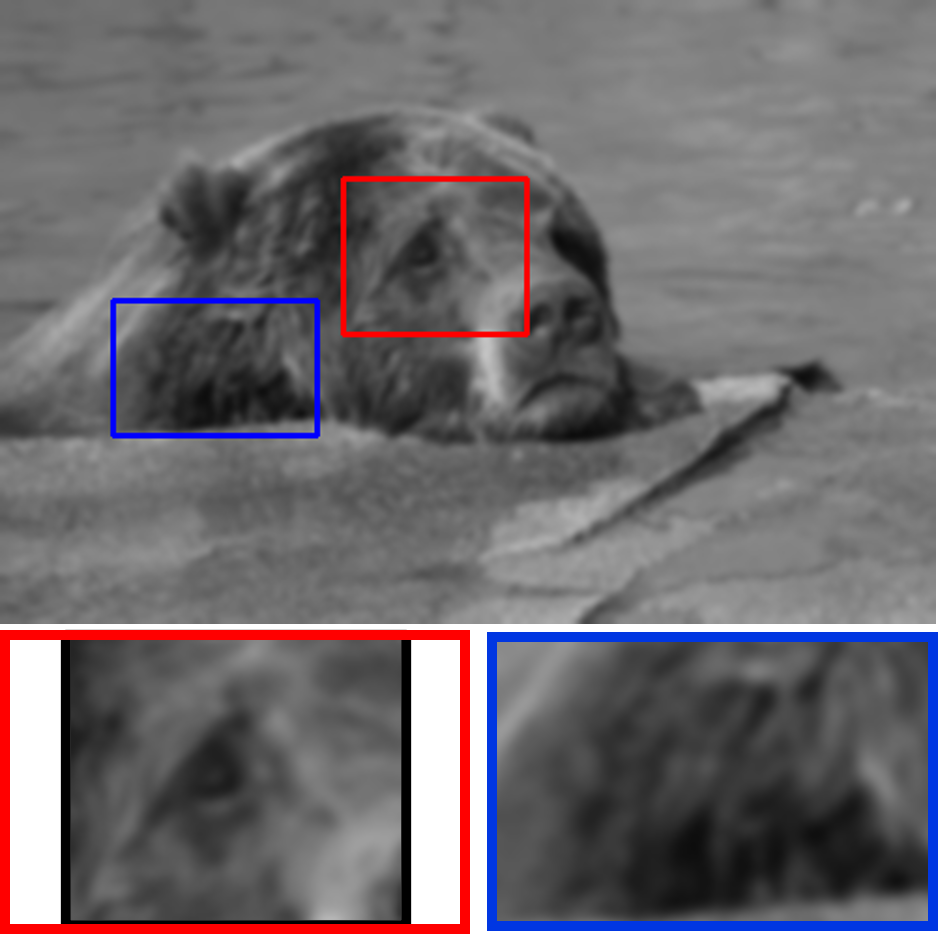}}    &
\subfigure[BME-SR (Scenario-1)]{\includegraphics[width=1.45 in]{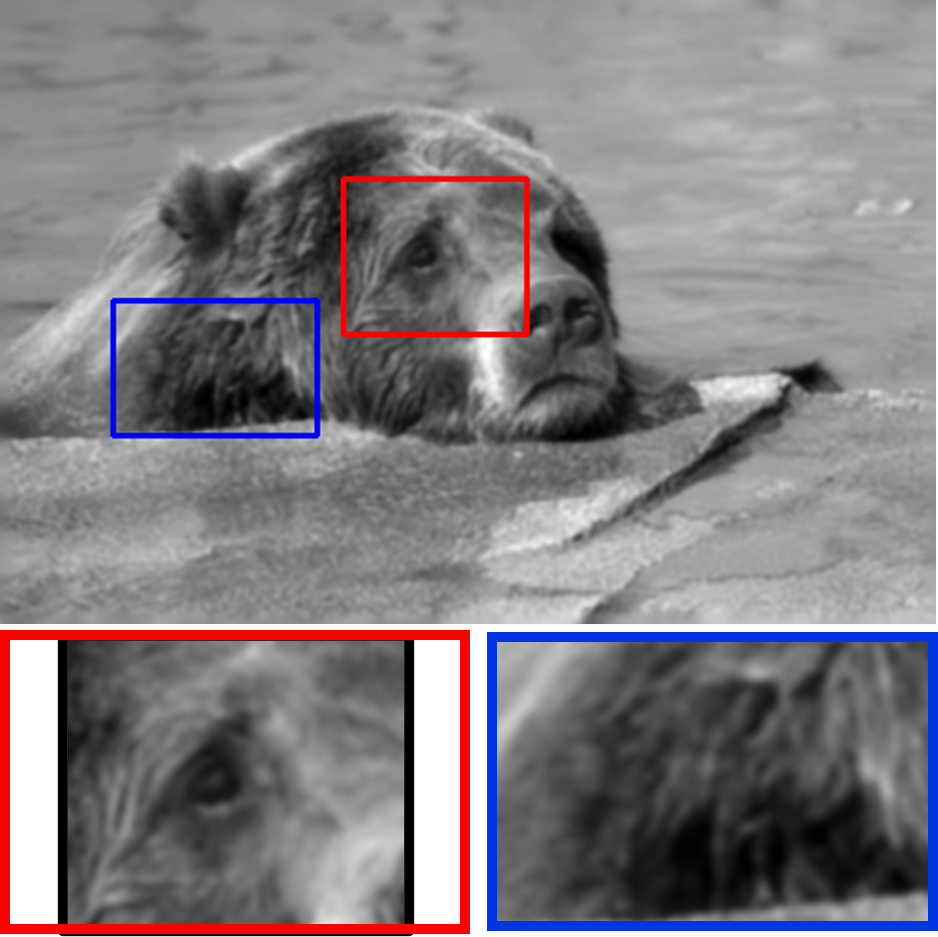}}    &
\subfigure[BME-SR (Scenario-2)]{\includegraphics[width=1.45 in]{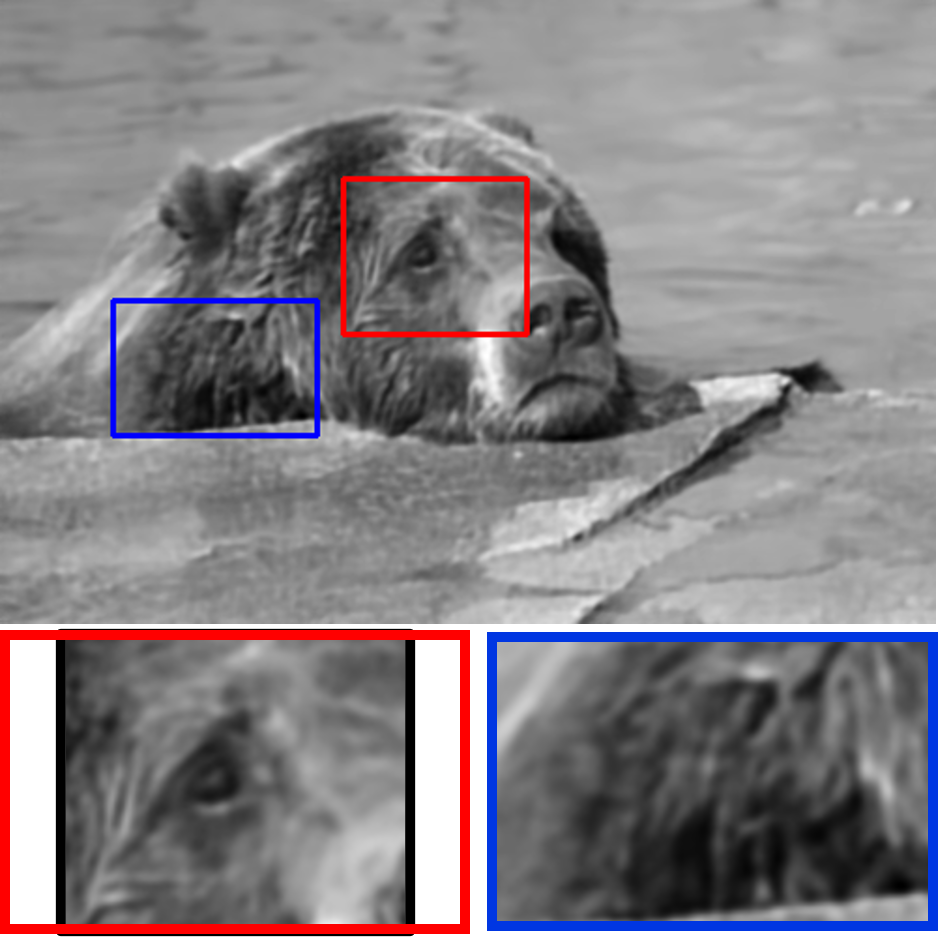}} \\
\subfigure[HR]{\includegraphics[width=1.45 in]{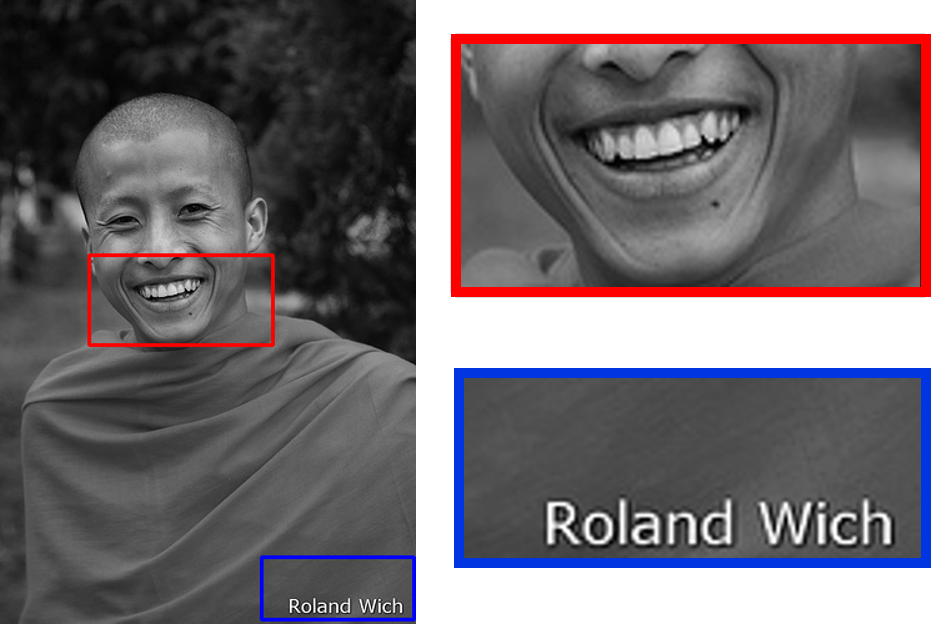}} &
\subfigure[LR]{\includegraphics[width=1.45 in]{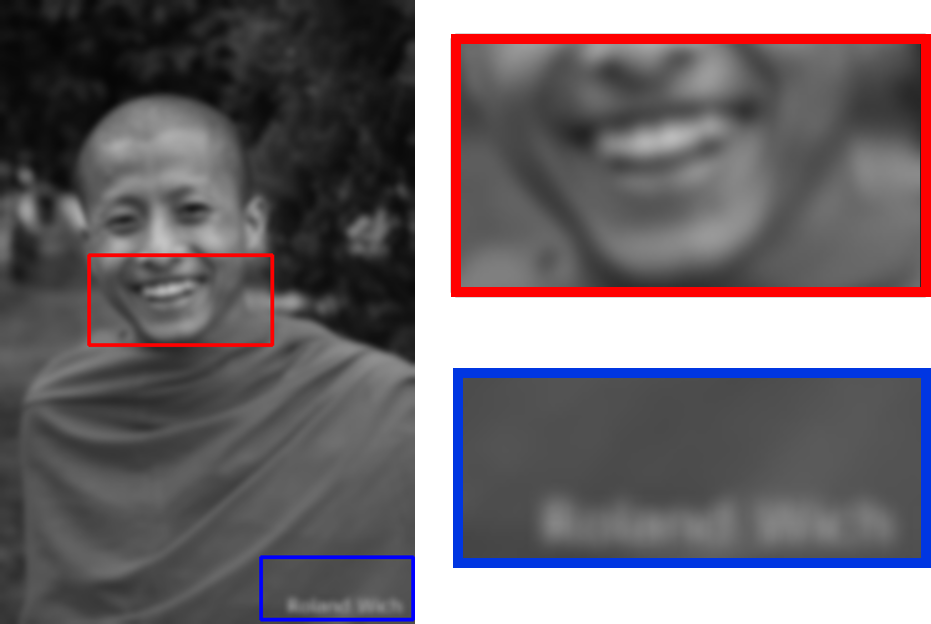}}  &
\subfigure[BME-SR (Scenario-1)]{\includegraphics[width=1.45 in]{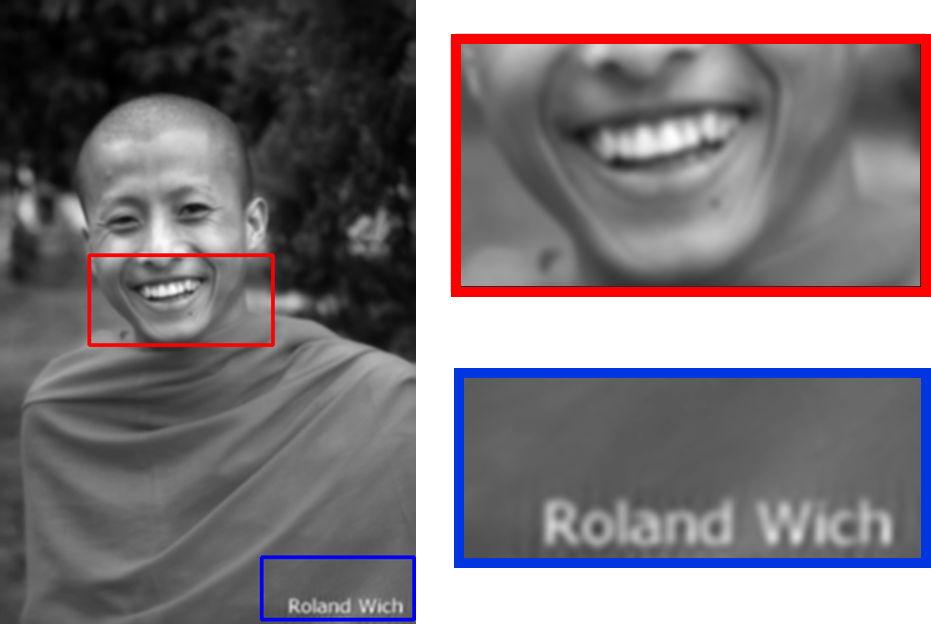}}  &
\subfigure[BME-SR (Scenario-2)]{\includegraphics[width=1.45 in]{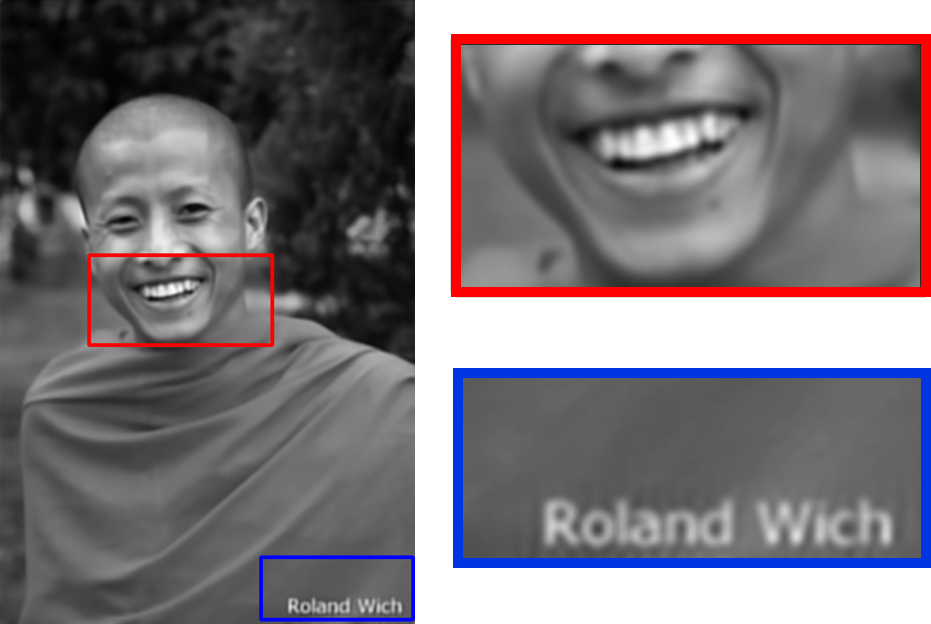}}
\end{tabular}
\caption{Results for two scenarios of non-correspondence data on the Gaussian dataset. $I_{5}$ (first row, blur: $k=7$, $\sigma=1.2$), $I_{6}$ (second row, blur: $k=11$, $\sigma=2.5$), and $I_{7}$ (third row, blur: $k=9$, $\sigma=2.0$). Columns show: (a, e, i) HR ground truth, (b, f, j) LR input, (c, g, k) scenario-1 reconstruction, and (d, h, l) scenario-2 reconstruction.}
\label{fig:UP-NC_result}
\end{center}
\end{figure*}
For solving \eqref{eq:NC-dictionary}, we adapt the standard K-SVD algorithm to handle the joint optimization problem. We employ an atom-wise optimization strategy where each dictionary atom is updated iteratively while keeping all other parameters fixed. This approach extends the classical K-SVD framework to accommodate dual-domain constraints arising from unpaired data \cite{kSVD}.

For each dictionary atom $\mathbf{D}^h_{:,t}$, we first identify the support sets 
\begin{align}
\omega_t &= \{i : \mathbf{C}_{t,i} \neq 0\}, \text{ and } \widetilde{\omega}_t = \{j : \widetilde{\mathbf{C}}_{t,j} \neq 0\},
\label{eq:support-sets}
\end{align}
which represent the indices of samples that utilize the $t$-th atom in the LR and HR domains, respectively. Where, $\mathbf{D}^{h}_{:,t}$ is the t-th column of HR dictionary $\mathbf{D}^{h}$, while $\mathbf{C}_{t,i}$ and $\widetilde{\mathbf{C}}_{t,j}$ are the corresponding $\left( t,i \right)$ and $\left( t,j \right)-th$ sparse coefficient of $\mathbf{C}$ and $\widetilde{\mathbf{C}}$ respectively. The current atom's contribution is then temporarily removed by setting $\mathbf{D}^h_{:,t} = \mathbf{0}$, allowing us to compute the reconstruction residuals for both domains.

The residual computation involves calculating the HR domain error as 
\begin{align}
\mathbf{E}^h_t &= \mathbf{X}^h_{:,\widetilde{\omega}_t} - \mathbf{D}^h \widetilde{\mathbf{C}}_{:,\widetilde{\omega}_t},
\label{eq:hr-residual}
\end{align}
and the LR domain error as 
\begin{align}
\mathbf{E}^l_t &= (\mathbf{B}^T\mathbf{B})^{-1}\mathbf{B}^T(\mathbf{Y}^l_{:,\omega_t} - \mathbf{B}\mathbf{D}^h\mathbf{C}_{:,\omega_t}).
\label{eq:lr-residual}
\end{align}
These residuals quantify the representation gaps when the $t$-th atom's contribution is excluded from the reconstruction.

The key innovation lies in the joint optimization setup, where we concatenate the residual matrices and the corresponding coefficient vectors as
\begin{align}
\mathbf{E}_t = [\mathbf{E}^h_t \mid \mathbf{E}^l_t], \quad 
\boldsymbol{\gamma}_t = [\widetilde{\mathbf{C}}_{t,\widetilde{\omega}_t} \mid \mathbf{C}_{t,\omega_t}]. \label{eq:concat-coeff}
\end{align}
This concatenation transforms the dual-domain optimization problem into a unified rank-1 approximation task
\begin{align}
\min_{\mathbf{d}^h_t, \boldsymbol{\gamma}_t} \left\| \mathbf{E}_t - \mathbf{D}^{h}_{:,t} \boldsymbol{\gamma}_t \right\|_{F}^{2},
\label{eq:rank1-approx}
\end{align}
enabling simultaneous minimization of both reconstruction errors.

The atom update is performed using SVD decomposition of the concatenated residual matrix $\mathbf{E}_t$. Given the SVD decomposition
\begin{align}
\mathbf{E}_t &= \mathbf{U} \boldsymbol{\Sigma} \mathbf{V}^T,
\label{eq:svd-decomp}
\end{align}
the optimal dictionary atom and coefficients are obtained as
\begin{align}
\mathbf{D}^h_{:,t} = \mathbf{U}_{:,1}, \text{ and } \boldsymbol{\gamma}_t = \boldsymbol{\Sigma}_{1,1} \mathbf{V}^T_{1,:}, \label{eq:dic-coeff-update}
\end{align}
where $\mathbf{U}_{:,1}, \mathbf{V}_{1,:}$, and $\boldsymbol{\Sigma}_{1,1}$ are the first singular vectors and singular value, respectively. The corresponding sparse coefficients are then decomposed back into the respective domain-specific coefficient vectors.

This atom-wise approach preserves the computational efficiency of K-SVD while satisfying the dual-domain constraints in Eq.~\eqref{eq:NC-dictionary}. The method leverages the shared dictionary structure to enforce feature consistency across HR-LR domains, ensuring each atom captures degradation-invariant patterns despite the absence of explicit correspondences. The complete procedure is summarized in Algorithm~\ref{alg:joint-dict-update}.

The absence of explicit patch-level correspondences between LR and HR domains significantly complicates joint estimation of $\mathbf{B}$ and $\mathbf{D}^{h}$. The atom-wise dictionary update with integrated sparse coding leverages statistical correlations across domains. This dual regularisation strategy, structural for the blur matrix and sparse for representations, enables robust optimisation despite incomplete supervision. 

Following joint estimation of $\mathbf{D}^{h}$ and $\mathbf{B}$, the inference procedure remains consistent with the correspondence-based framework. LR patches are decomposed via $\mathbf{D}^{l} = \mathbf{B}\mathbf{D}^{h}$, with sparse coding and subsequent HR reconstruction performed identically to Section~\ref{sec:Paired_data}. The next sections present the results of implementing the suggested approach on unpaired and non-corresponding datasets.

\subsection{Experiments and Results}
In this experiment, we utilized the same four subsets from the CMU-Cornell iCoseg database~\cite{iCoseg-database}, where 7 images were selected for training and the remaining images were used for testing, as described previously. For each training subset, we randomly extracted $20{,}000$ image patches of size $15 \times 15$ pixels. Unlike the previous setup, where each LR patch had a corresponding HR counterpart, we now considered two distinct non-correspondence scenarios regarding the relationship between LR and HR patches:

\begin{figure*}[!t]
\begin{center}
\begin{tabular}{ccccc}
\subfigure[LR]{\includegraphics[width=0.85 in]{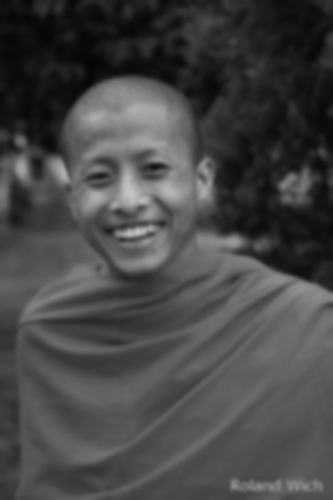}}   \quad &
\subfigure[$k = 7$]{\includegraphics[width=0.85 in]{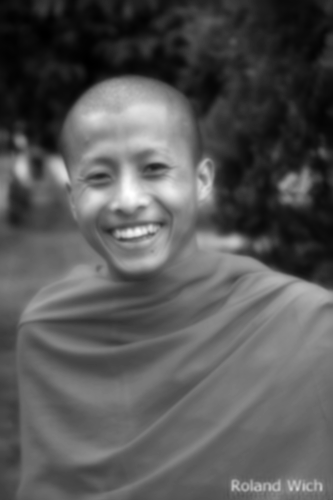}} \quad &
\subfigure[\textbf{$k = 9$}]{\includegraphics[width=0.85 in]{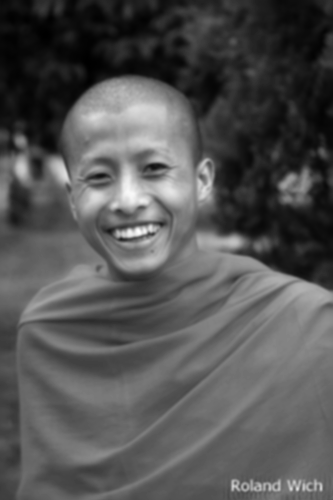}} \quad  &
\subfigure[$k = 11$]{\includegraphics[width=0.85 in]{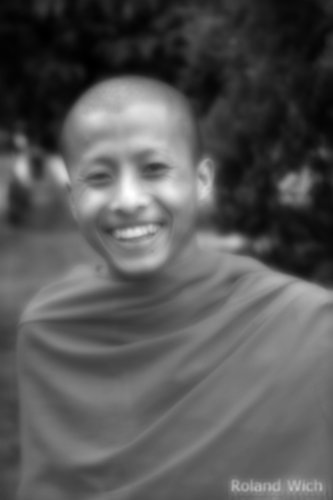}} \quad &
\subfigure[$k = 13$]{\includegraphics[width=0.85 in]{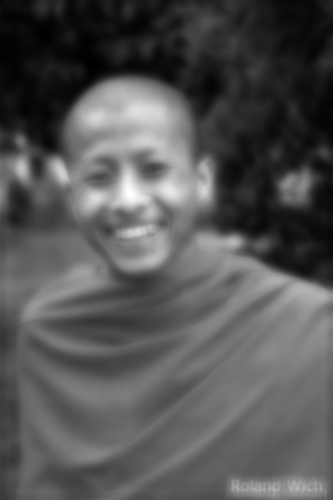}} 
\end{tabular}
\caption{Cross-validation results for scenario-2 on $I_{6}$. (a) LR input; (b-e) reconstructions with $k = 7, 9, 11, 13$ respectively}
\label{fig:Result_nc_cross_val}
\end{center}
\end{figure*}

\begin{enumerate}
\item \textbf{Scenario 1: Unpaired Data} \\
In this setting, there was no correspondence between the LR and HR patch sets. Specifically, the $20{,}000$ HR patches were sampled independently and were entirely different from the $20{,}000$ LR patches. There was no overlap or mapping between the two sets.
\item \textbf{Scenario 2: No-correspondence Data} \\
Here, the same set of $20{,}000$ spatial locations was used to extract both LR and HR patches; however, the order of patches in the LR and HR sets was randomly permuted, resulting in a lack of one-to-one correspondence. Thus, while the content was the same, the mapping between LR and HR patches was not preserved.
\end{enumerate}

\begin{table}[!t]
\centering
\caption{Scenario 1: Laplace and Sobel variance for different $k$ values} \small
\begin{tabular}{|c|c|c|c|c|c|}
\hline
Image & Metric & \multicolumn{4}{c|}{$k$ values} \\
\cline{3-6}
& & $k=13$ & $k=11$ & $k=9$ & $k=7$ \\
\hline
\multirow{2}{*}{$I_{5}$} & Sobel Var & 2945 & 2933 & 3102 & \textbf{3265} \\
& Laplace Var & 840 & 840 & 880 & \textbf{910} \\
\hline
\multirow{2}{*}{$I_{6}$} & Sobel Var & 2012 & 2122 & \textbf{2292} & 2281 \\
& Laplace Var & 510 & 540 & \textbf{550} & 540 \\
\hline
\multirow{2}{*}{$I_{7}$} & Sobel Var & 2684 & \textbf{2793} & 2612 & 2593 \\
& Laplace Var & 830 & \textbf{860} & 820 & 800 \\
\hline
\multirow{2}{*}{$I_{8}$} & Sobel Var & 1550 & \textbf{1620} & 1580 & 1510 \\
& Laplace Var & 450 & 480 & \textbf{500} & 460 \\
\hline
\end{tabular}
\label{tab:metrics_unpaired}
\end{table}

\begin{table}[!t]
\centering
\caption{Scenario 2: Laplace and Sobel variance for different $k$ values} \small
\begin{tabular}{|c|c|c|c|c|c|}
\hline
Image & Metric & \multicolumn{4}{c|}{$k$ values} \\
\cline{3-6}
& & $k=13$ & $k=11$ & $k=9$ & $k=7$ \\
\hline
\multirow{2}{*}{$I_{5}$} & Sobel Var & 2951 & 2927 & 3110 & \textbf{3268} \\
& Laplace Var & 845 & 836 & 877 & \textbf{915} \\
\hline
\multirow{2}{*}{$I_{6}$} & Sobel Var & 2007 & 2130 & \textbf{2287} & 2275 \\
& Laplace Var & 512 & 538 & \textbf{553} & 537 \\
\hline
\multirow{2}{*}{$I_{7}$} & Sobel Var & 2679 & \textbf{2794} & 2617 & 2599 \\
& Laplace Var & 828 & \textbf{859} & 818 & 803 \\
\hline
\multirow{2}{*}{$I_{8}$} & Sobel Var & 1556 & 1584 & \textbf{1617} & 1513 \\
& Laplace Var & 453 & 482 & \textbf{497} & 462 \\
\hline
\end{tabular}
\label{tab:metrics_nc}
\end{table}

These distinct scenarios were crucial for a comprehensive analysis because they mimic real-world situations where perfect correspondence between low- and high-resolution images is often unattainable. By evaluating our method under both scenarios, we could better assess its robustness and applicability in practical scenarios where data alignment is imperfect or entirely absent.

For both scenarios, we applied the proposed BME-SR method, which leverages the structural properties of Gaussian blur to estimate the blur matrix and to learn the high-resolution dictionary.

To optimize the performance of our approach, we conducted an exhaustive hyperparameter search for the FISTA algorithm, as well as for the learning rate of the descent algorithm, following the procedure outlined in the correspondence-based experiments. In cases where the sparsity parameter $k$ was unknown, and since traditional image quality metrics such as PSNR and SSIM could not be employed due to the absence of one-to-one correspondence between LR and HR images, we instead evaluated the results using the variance of Sobel and Laplacian edge responses. For each value of $k$, we computed the Sobel and Laplacian variances of the reconstructed images and selected the image with the highest variance as the final output, as higher variance typically indicates sharper and more detailed reconstructions. 

For the non-correspondence setting, we defined the test images as $I_{5}$, $I_{6}$, and $I_{7}$, corresponding to the HR ground truth images shown in Fig.~\ref{fig:UP-NC_result} panels (a), (e), and (i), respectively and their blurred counterparts
in panels (b), (f), and (j), respectively. A qualitative comparison is demonstrated using Figure~\ref{fig:Result_nc_cross_val}, showing deblurred images of $I_{6}$ for different $k$ in the unpaired setting. With $k = 9$ producing the best result, quantitatively supported by Sobel and Laplacian variance values from Table~\ref{tab:metrics_nc}.
Figure~\ref{fig:UP-NC_result} illustrates the results obtained for different $k$ and $\sigma$ values in both scenarios. From the second example of $I_{6}$, we could clearly retrieve the blurred text from LR in our results; also, the teeth were more clearly visible than in the LR image. For each case, images with the best cross-validation result are shown in the comparison for both scenarios. The CDL method could not be used in this setting of unpaired data, and all other existing algorithms failed under this scenario where there was no one-to-one correspondence and low data availability. Tables~\ref{tab:metrics_unpaired} and \ref{tab:metrics_nc} report the Laplacian and Sobel variances for various images and $k$ values for Scenario 1 and 2, respectively. 

Also interestingly, our experiments revealed that the results obtained under both scenarios were quite similar. This is supported by the results in Tables~\ref{tab:metrics_unpaired} and \ref{tab:metrics_nc}, and Figure~\ref{fig:UP-NC_result}. This suggests that the performance of our method is not heavily dependent on even a weak form of correspondence between LR and HR patches. We can infer that our approach effectively leverages the underlying statistical properties of the image data, rather than relying on specific spatial relationships between corresponding patches, making it highly adaptable to situations with limited or no alignment information.


\section{Conclusion}
\label{sec:conclusions}

In this research, we presented a novel dictionary learning based deblurring framework that successfully addresses the important problem of image deblurring with minimal data. Our approach enables reliable model training regardless of diverse levels of data supervision. By jointly estimating a structured blur pattern and a detailed image dictionary, this method fundamentally reduces the need for large paired datasets. Experiments on the CMU-Cornell iCoseg and FocusPath datasets in paired, unpaired, and non-correspondence data conditions repeatedly demonstrated that this approach outperforms current techniques in deblurring. These findings confirm that accurate blur characterization and adaptive dictionary construction are achievable with very limited training data. This makes our method extremely useful for domains like medical imaging and remote sensing, where it is challenging to generate completely matched datasets. In order to combine interpretability and learning capacity, future work may incorporate deep unrolling techniques and extend the model to handle spatially varied and nonlinear blur models.



\bibliographystyle{IEEEtran}
\bibliography{references}





\end{document}